\documentclass[fleqn,usenatbib]{mnras}

\usepackage{newtxtext,newtxmath}

\usepackage[T1]{fontenc}

\DeclareRobustCommand{\VAN}[3]{#2}
\let\VANthebibliography\thebibliography
\def\thebibliography{\DeclareRobustCommand{\VAN}[3]{##3}\VANthebibliography}

\usepackage{graphicx}	
\usepackage{amsmath}	
\DeclareUnicodeCharacter{2212}{-}
\usepackage{caption}
\usepackage[frozencache,cachedir=.]{minted}

\title[Transferability of Photometric Redshifts Determined using Machine Learning]{Testing the Transferability of Machine Learning Techniques for Determining Photometric Redshifts of Galaxy Catalogue Populations}

\author[Lara Janiurek et al.]{
Lara Janiurek,$^{1}$\thanks{E-mail: lara.janiurek@strath.ac.uk}\thanks{Current institution: University of Strathclyde, John Anderson Building, 107 Rottenrow E, Glasgow G4 0NG}
Martin A. Hendry,$^{1}$
Fiona C. Speirits$^{1}$
\\
$^{1}$SUPA, School of Physics and Astronomy, University of Glasgow, University Avenue, Glasgow, Scotland G12 8QQ\\
}

\pubyear{2023}

\begin{document}
\label{firstpage}
\pagerange{\pageref{firstpage}--\pageref{lastpage}}
\maketitle

\begin{abstract} Galaxy redshift surveys are an essential component of modern cosmological research. However, photometric redshift surveys are often subject to significant uncertainties and spectroscopic surveys are resource-intensive and expensive to obtain. Developing and applying methods to generate more accurate photometric redshift estimates would, therefore, be very beneficial for cosmological inference. In this work, the random forest algorithm GALPRO is implemented to generate photometric redshift posteriors, and its performance when trained and then applied to data from another survey is investigated and assessed. The algorithm is initially calibrated using a truth dataset compiled from the DESI Legacy survey. Tests are run using the DESI dataset to determine how statistically similar the training and testing datasets from this survey must be in order that the results of applying GALPRO to the training data are also fully applicable to the testing data. We find that the testing and training datasets must have very similar redshift distributions, with the range of their photometric data overlapping by at least 90\% in the appropriate photometric bands, in order for the results obtained for the training data to be applicable to the testing data.

In a further test, GALPRO is again trained using the DESI dataset and then applied to a sample drawn from the PanSTARRS survey, to explore whether GALPRO can be first trained using a trusted dataset and then applied to an entirely new survey -- albeit one that uses a different magnitude system for its photometric bands, thus requiring careful conversion of the measured magnitudes for the new survey before GALPRO can be applied. The results of this further test indicate that GALPRO does not produce accurate photometric redshift posteriors for the new survey, even where the distribution of redshifts for the two datasets overlaps by over 90\%.  Hence, we conclude that the photometric redshifts generated by GALPRO are not generally suitable for generating estimates of photometric redshifts and their posterior distribution functions when applied to an entirely new survey, particularly one that uses a different magnitude system. However, our results demonstrate that GALPRO is a useful tool for inferring photometric redshift estimates in the case where a spectroscopic galaxy survey is {\em nearly\/} complete, but is missing some spectroscopic redshift values.
\end{abstract}




\section{Introduction}
\label{intro}

The use of galaxy catalogues as cosmological probes is an important and widely-studied field.  However, before being able to infer cosmological parameters from a galaxy catalogue, the redshift of each galaxy must first be obtained. Cosmological redshift occurs as photons emitted from a distant source travel through the universe and the light emitted from the source is shifted to longer wavelengths as it traverses expanding spacetime \citep{particles7020019}. This redshift is therefore primarily due to the expansion of space itself and is not caused by the motion of an object or observer -- although a component of the redshift may be due to these latter factors, particularly for nearby galaxies \citep{space}. The further the object is from the observer, the higher the redshift and the greater its recessional velocity, as described by the Hubble Law. 
The redshift of extragalactic objects is a difficult property to measure, however, and may be determined from spectroscopic or photometric observations.

Spectroscopic redshifts are subject to much smaller random and systematic errors than photometric techniques; however, they are also much more resource-intensive (e.g. in terms of telescope time) and therefore often more difficult to measure \citep{Quadri_2010}. In view of the challenges associated with acquiring large numbers of spectroscopic redshifts, over the past few decades there has been significant and increasing attention focused on improving photometric redshift methods, which can in principle be measured more efficiently and quickly at scale \citep{jones2023photometric,red2}. Photometric redshift techniques infer galaxy redshifts by measuring the brightness of a galaxy observed through broadband photometric filters and make use of information about how features in the galaxy spectra move through those filters, thus changing the relative brightness of the galaxy in the different bands, as the distance (and hence cosmological redshift) of the galaxy increases \citep{red45}. 

Photometric redshifts are advantageous in that they allow for the efficient estimation of cosmological redshifts for large numbers of galaxies identified in an imaging survey. However, the precision of photometric redshift inference is typically at least about a factor of ten worse, compared with the precision of the redshift of a galaxy measured with a low-resolution spectrograph \citep{flavours,Zhou_2021,red2,2024MNRAS.530.2012T}. This lack of precision is in part due to the fact that the photometric filters capture all the information in a galaxy spectrum across a broad wavelength range and reduce that information to a single flux measurement. Hence, for any given galaxy, the relationship between the template spectrum used and the nature and detailed features of the {\em true} galaxy spectrum with which it is being compared is only approximately modelled.

There are two main techniques for determining photometric redshifts. The first of these, template fitting approaches \citep{LYF, Brammer_2008, 10.1093/mnras/stx2536, Steinhardt_2023}, utilise models of the spectral energy distribution (SED) of a galaxy to approximate the likelihood of a galaxy's observed photometry, conditional upon its redshift. Bayesian approaches, such as \citet{photoz1,10.1093/mnras/sty3279,Leistedt_2023}, then combine these likelihoods with prior information to infer a posterior probability density function (PDF) of the redshift given a galaxy's observed photometry.

An important alternative to template fitting methods is to make direct use of {\em observed} broadband photometric data and measured spectroscopic redshifts for a set of galaxies to derive the statistical relationship between their photometric and spectroscopic data -- for example using principal component analysis to determine a linear relation between the two \citep{1995AJ....110.2655C}, such that a galaxy's spectroscopic redshift can then be predicted from its measured photometric data.

An interesting, and increasingly popular, sub-category of these data-driven approaches makes use of machine learning (ML) methods to estimate photometric redshifts -- see, for example \cite{ml6}; \cite{ml5}; \cite{ml3};  \cite{Norris_2019}; \cite{ml4}; \cite{ml2}; \cite{red2}; \cite{10.1093/mnras/stad3976} and references therein. In such approaches, redshifts are again inferred without making any use of template spectra; instead a ``training sample'' of galaxies is again identified that has both a measured spectroscopic redshift and photometric data for each galaxy. These training data are then processed by some form of ML algorithm that ``learns'' the relationship between the galaxies' measured spectroscopic redshifts and photometric fluxes -- so that the learned relationship may then in principle be applied to another galaxy to estimate its redshift from its measured broadband photometry. 

Since these ML approaches do not rely on model spectra, in principle they should avoid any bias or artificial structures that may be introduced by such models \citep{photoz1,2010ApJ...712..511C} -- although they may still be vulnerable to biases arising from incompleteness in the spectroscopic training samples used to train them.  More generally, the relationship learned by the ML algorithm may be sensitive to the details of the galaxy training data used or the particular algorithm adopted, which may then limit the applicability of the algorithm to new datasets outside the range of the training sample \citep{10.1093/mnras/stab1513,red2,10.1093/mnras/stad3976}.
Several different ML-based methods for estimating photometric redshifts have been studied in recent literature, including random forest (RF) techniques \citep{2013MNRAS.432.1483C}, feed-forward neural networks \citep{2013ApJ...772..140B}, Gaussian processes \citep{2016MNRAS.462..726A}, boosted decision trees \citep{ml3} and support vector machines \citep{2017A&A...600A.113J}. 
Whether template-based or data-driven methods are employed to estimate photometric redshifts, however, it seems clear that thoroughly assessing the accuracy and robustness of the methodologies adopted is important before they are used for cosmological inference. For the ML-based methods mentioned above, for example, such an assessment and ``benchmarking" of the different methodologies against each other has been enabled by the PHoto-z Accuracy Testing (PHAT) program \citep{2010A&A...523A..31H}.  Currently, however, no single photometric redshift estimation method appears to be superior in all circumstances to all of the others \citep{refId0,red2}.

The purpose of this paper is to assess the accuracy and robustness of one such example of an ML-based methodology: the GALPRO method \citep{2021MNRAS.502.2770M} that uses a random forest (RF) approach to infer photometric redshift estimates. 
 We will focus in particular on the applicablity of GALPRO to new testing data that overlaps only partially with the distribution of the data used to train the RF -- an issue that, as noted e.g. in the recent review by \cite{red2}, can result in limitations in the performance of method.

In recent years many spectroscopic and photometric redshift surveys have been used to carry out such ``benchmarking" of photometric redshift estimation methods. These surveys include \citet{LYF}, \citet{LY}, \citet{Sanchez} and \citet{10.1093/mnras/stx687}. Initial data-driven photometric redshift methods provided point estimates, while \citet{num1} employed a neural network to estimate photometric redshift PDFs with realisations scattered within photometric errors. \citet{wolf} later used a data-driven approach to estimate redshift PDFs, highlighting the intrinsic floor in redshift uncertainties, even under ideal conditions, resulting from the inherent redshift spread in the training set. 

As was previously noted, ML methods, which map the photometric data space to the redshift estimate, come with their own set of limitations due to e.g. incompleteness in the training data. The quality of the photometric redshift estimates, therefore, depends on a variety of factors such as the uniformity of coverage of the observable parameter space, the variety and depth of photometric information and the fraction of peculiar objects contained in the survey~\citep{Brescia_2021,red2}.

Various quantitative studies of the robustness of ML-based methods have been carried out in recent literature. For example, \cite{10.1093/mnras/stab1513} investigated deep learning methods for inferring photometric redshift estimates directly from pixel-level information contained in galaxy images, compared these with more conventional ML methods based on multi-band photometry, and use various metrics including the bias, precision and outliers fraction (see Section 
\ref{sec:performance}) to assess their performance. Their results indicate that image-based deep learning methods generally require larger training samples to reach their optimal performance and also highlight the importance of the training data being capable of predicting redshifts across the full range encountered in the testing data. Similar conclusions are reached in \cite{2022MNRAS.515.5285D}, which uses a deep learning network trained with multi-band images, spectroscopic redshifts, and Galaxy Zoo classifications of around 400,000 Sloan Digital Sky Survey (SDSS) galaxies. However, \cite{red2} note that such image-based methods will likely yield less impressive results when applied to fainter, poorly-resolved galaxies in future deep imaging surveys for which only limited spectroscopic redshift training data exists.

In another recent study, \cite{10.1093/mnras/stad3976} have used a weighted RF algorithm to study the effectiveness of photo-$z$ estimation methods for the multi-band survey planned by the Chinese Space Station Telescope (CSST), using a training set of simulated data adapted from the Hubble Space Telescope Advanced Camera for Surveys and COSMOS catalogues.  These authors show that their weighted RF method generally produces accurate results, but its performance is less impressive for obtaining photo-$z$ estimates of {\em low\/} redshift galaxies, due primarily to there being relatively few training samples in this redshift range.

In this work we build upon the example of \cite{10.1093/mnras/stad3976} to further investigate the robustness of RF methods, utilising the software package GALPRO \citep{galpro-github} to generate photometric redshifts. We aim to determine whether GALPRO, trained and calibrated on a given survey, is then fully {\em transferable\/}, i.e.~is applicable to a new survey for which only photometric data are available, in order to allow generation of reliable and accurate photometric redshifts for galaxies where no spectroscopic data are available. 
GALPRO is an RF algorithm that can be trained using observed photometry and spectroscopic redshifts, to learn the mapping between the two. Once the mapping is learnt, it can then in principle be applied to another, different galaxy survey where there is a lack of spectroscopic data, in order to generate photometric redshifts for those galaxies. This work explores how reliable the photometric redshift estimates may be when applying the GALPRO software to a different survey in this manner. By applying GALPRO to a new, different survey which is completely separate from the one on which it was trained, we will then seek to evaluate the accuracy of the results obtained.

We note that in this paper (including in its title) we refer to this property as the `transferability' of the GALPRO results; however, we should clarify that this terminology does {\em not\/} relate in any way to the specific concept of `transfer learning' in machine-learning studies -- see, for example, \cite{Bozinovski2020ReminderOT}. In fact, our use of the term `transferability' is closer in meaning to the concept of `covariate shift' in multivariate analysis and machine-learning studies \citep{8978471}, i.e.~where the distribution of the independent variables changes between the training and testing environments.

The remainder of this paper is organised as follows. Section \ref{sec_2} introduces the GALPRO software package and details its calibration and validation using known data samples. Section \ref{section_3} then assesses GALPRO's performance when applied to a new survey which may contain galaxies with different properties than those on which it was trained. Specifically, Section \ref{overlap}
first considers the case where the distributions of photometric fluxes in the calibration data and validation data partially overlap, and considers the impact of varying the degree of overlap between the two sets of data. Next, Section \ref{pan_section} explores the transferability of GALPRO to a completely different survey, for which in principle only photometric data might exist. In each case, the results of using GALPRO are evaluated to determine the efficacy of the software when applied to a (partially or completely) new survey to generate photometric redshifts.  Finally, Section \ref{summary_conc} presents a summary of our results and conclusions.

\section{GALPRO Calibration}\label{sec_2}

\subsection{Random Forest Algorithms} 

The core principles of photometric redshift estimation lie in determining the mapping between a range of colours (or fluxes) and redshift \citep{82}. Comparing this mapping with the observed fluxes of a source under study then yields a point estimate for the source's redshift -- or, better, allows inference of a probability distribution function (PDF) from which the redshift may then be estimated \citep{red2}. As already discussed in Section \ref{intro}, this mapping may be determined by several methods, including using ML approaches, where a representative training sample with known redshifts and photometry is employed, or using template-fitting methods, whereby the redshift-colours mapping is based upon theoretical models or previous scientific knowledge \citep{rfforphotoz}. The use of additional prior information can improve the performance of both approaches \citep{Tanaka_2015,red2}. Supervised ML demands both spectroscopic redshifts and photometry as training data, so the quality of the training data that is available dictates the precision of the output.

In this work we investigate a recent example of an ML algorithm used to infer photometric redshifts, known as GALPRO \citep{galpro-github}. GALPRO is a Python software package, developed in 2020, which may be used to compute galaxy properties such as redshift, star formation rate, stellar mass, etc. As noted by the authors, particularly useful is the ability of GALPRO to generate -- `on the fly', reasonably quickly and for large numbers of galaxies -- multivariate posterior probability distributions for these properties, and the case of 2-dimensional posteriors for redshift and stellar mass is the main focus considered in \citet{galpro-github}.  In our work we use GALPRO as an illustrative and instructive example with which to investigate the transferability of an ML algorithm from one survey to another\footnote{Clearly, other ML algorithms exist in the literature and the results presented in this paper investigating transferability might differ somewhat in quantitative detail in the case of other algorithms -- although we expect that our general, qualitative conclusions about the possible limits of transferability will apply more widely.}. 

GALPRO utilises an RF algorithm to form a supervised ML program \citep{galpro-github}. RFs may be used for the regression or classification of large datasets, and here GALPRO is used for regression purposes to compute the photometric redshift of galaxies.

In the case of estimating photometric redshifts, the RF algorithm determines a function which maps between the spectroscopic redshift values of the training sample and their multi-dimensional photometry space, i.e.~the fluxes and colours of each galaxy. It samples a random subset of the training data, which comprises fluxes, colours and spectroscopic redshifts, to build the decision trees \citep{b1}. Every cell in the decision tree is then successively assigned an average spectroscopic redshift value. A photometric redshift estimate may then be made by inputting the photometric properties of a galaxy, in this case the flux and colours, and passing these properties through each tree. An average is then taken over all of the obtained redshifts to produce the final photometric redshift estimation of the individual galaxy \citep{rftour}. A detailed description of how GALPRO generates redshift estimates and PDFs using random forests can be found in the documentation given in \citet{galpro-github}.

A photometric redshift posterior is much more informative than an individual point estimate \citep{Wittman_2009}. The posterior is a probability distribution characterising the probability that the galaxy really is at a given redshift, and so the shape of the posterior directly relates to the shape of the error distribution on redshift \citep{Brescia_2021}. GALPRO is also able to generate multivariate posterior distribution functions for its individual galaxy estimates. This is implemented by using the RF algorithm to predict both the photometric redshift and rest frame $r$-band absolute magnitude simultaneously. As this work is primarily concerned with inferring the population of redshift estimates for a galaxy survey, this multivariate feature is not relevant to the current study and will not be discussed further. Clearly, however, it may be useful in other contexts and (as already noted) is one of the key motivating features of the GALPRO approach. 

\subsection{Redshift Truth Table}\label{hyper}

GALPRO has previously been used to compute photometric redshifts for the DESI Legacy survey -- see \cite{zhou-desi}. Since this sample has already been successfully used with GALPRO, it is an obvious choice for our initial calibration. The RF algorithm used by GALPRO takes as input features the $r$-band magnitude and $r$-$z$, $g$−$r$, $z$-$W1$ and $W1$-$W2$ colours. The Zhou et al. sample gives the photometric band data in fluxes which can then be converted to {\em luptitudes\/} -- i.e.~the informal name used for the magnitude system adopted within the Sloan Digital Sky Survey (SDSS). These magnitudes are related to the corresponding photometric fluxes via an inverse hyperbolic sine function, as opposed to a logarithmic function.  This modified transformation is designed to be almost identical to the standard logarithmic relationship for high signal-to-noise ratio observations, but produces well-behaved magnitudes and colours even in the case of low signal-to-noise ratios. See \cite{1999AJ....118.1406L} for further details\footnote{See also https://www.sdss3.org/dr10/algorithms/magnitudes.php which gives a table of the numerical values for the conversion parameters relevant to the SDSS photometric bands.}. 

The Zhou et al. sample also takes the ratio between the semi-minor and semi-major axes, the half light radius, and a model weight as morphological parameters. However these parameters are omitted from our analysis as the PanSTARRS dataset considered in Section \ref{section_3} does not explicitly contain them. It was found that the omission of these parameters has no significant impact on the calibration and validity of the redshift PDFs \citep{me}.

All of the imaging catalogues used overlap with the DECaLS catalogue footprint and have been corrected for galactic extinction \citep{zhou-desi}. The majority of the galaxies in the `truth' dataset are taken from the SDSS, BOSS, GAMA and WiggleZ surveys, which are limited to shallow magnitudes and apply colour selections -- see \citet{100,boss,gama,wigglez} for more details. This causes sharp peaks in the redshift distribution which can introduce systematic bias as a result of the training sample being non-uniform, since the RF algorithm may favour redshift estimates that are over-represented in the truth dataset. To avoid this, galaxies from the four surveys are downsampled to create a more uniform training dataset. For more detailed information on the truth table and downsampling process, see \citet{zhou-desi}. 

To improve the performance of the algorithm, the RF hyperparameters may be tuned. Following results described in \cite{galpro-github}, which used the same truth sample as our work, the optimal hyperparameters were found and are shown in Table \ref{hyper_tab}.

\begin{table}
	\centering
	\caption{The random forest hyperparameter settings used throughout the analysis in this work, to improve the performance of the GALPRO.}
	\label{hyper_tab}
	\begin{tabular}{lr} 
		\hline
		Hyperparameter & Setting \\
		\hline
		n\_estimators & 2 \\
		max\_features & 4 \\
		max\_depth & 5 \\
        min\_samples\_leaf & 3 \\
        min\_samples\_split & 2 \\
        max\_leaf\_nodes & none \\
        min\_impurity\_decrease & 0.0 \\
        min\_impurity\_split & none \\
        min\_weight\_fraction\_leaf & 0.0 \\ 
		\hline
	\end{tabular}
\end{table}

This selection of hyperparameters allows the decision trees to be fully grown, meaning the training data may no longer be split. The choice of `max\_features' being `auto' guarantees that the RF algorithm has a sufficient amount of prior information, which is essential as we are only using a small number of photometric bands.

\subsection{Performance Assessment}
\label{sec:performance}

To thoroughly assess the performance of the GALPRO algorithm and the validity of the results that it produces, multiple measures can be derived from the redshift estimates and their PDFs. This is usually done by comparing for each galaxy in the validation sample the photometric redshift point estimate (eg. adopting the mean or mode of the PDF as $z_{\rm photo}$) and the spectroscopic redshift ($z_{\rm spec}$). Note that any spectroscopic redshift values used in ML training or prior construction should be discarded from the validation sample. 

Assessing the performance of photometric redshift point estimates usually entails measuring the following:

\begin{itemize}
  \item {\em Precision\/} ($\sigma_{\rm NMAD}$). This describes the 68$^{\rm th}$ percentile width of the distribution about the median, and is defined as either the standard deviation of ($z_{\rm phot} - z_{\rm spec}$)/(1 + $z_{\rm spec}$) or 1.48 $\times$ median($|z_{\rm phot} - z_{\rm spec}|$)/(1 + $z_{\rm spec}$) \citep{metric}.
  
  \item {\em Bias\/}. This describes the average separation between the true and predicted redshifts and is defined as $\langle z_{\rm phot} - z_{\rm spec} \rangle$ \citep{Dahlen_2013}.
  
  \item {\em Outliers fraction\/}. This gives the fraction of anomalous sources with unexpectedly large error values, defined by $|z_{\rm phot} - z_{\rm spec}|/(1+z_{\rm spec}) > 0.15$ \citep{flavours}. The value of 0.15 is a standard value used to define the catastrophic outlier fraction \citep{0.15_ref}.
\end{itemize}

The probability integral transform (PIT) is an extremely useful tool for validation, as it indicates if any bias has been introduced to the PDFs. The PIT is used to assess probabilistic calibration and is defined as the cumulative distribution function (CDF), evaluated at the true redshift \citep{pit15}. The PIT plot demonstrates how uniform the CDF accumulation is compared to an expected uniform PIT from 0 to 1, denoted by U$(0,1)$ \citep{90}. The PIT distribution will display a convex shape if the marginal PDFs are too broad, as less objects will have true redshifts within the tails of their PDF \citep{concave}. Conversely, if the PIT is a concave shape, the PDFs are overly narrow and too many objects contain the true redshift within the tails of their PDF. The PIT distribution must be uniform in order for the marginal PDFs to be considered valid, although even a uniform PIT may also contain some bias. 

There are many methods of qualitatively measuring the uniformity of the PIT, such as the Kullback–Leibler divergence (KLD), Kolmogorov–Smirnov test (KST) and Cramer-von Mises (CvM) tests \citep{97,98,99}. These metric tests all determine the similarity between the PIT and U$(0,1)$. All of these tests result in a value of zero if there is a perfect match between the two distributions. The larger the absolute value of each test, the larger the deviation of the PIT from a uniform distribution.

The equality of the true and predicted redshift distributions may be assessed using 
{\em marginal calibration\/}. To accurately describe the redshift PDFs as marginally calibrated, the average predictive and true empirical PDFs must be completely or almost equal. The marginal calibration plots shown in later sections of this work represent the difference between the average predictive and true empirical PDFs at regular intervals \citep{marg_cal}. If this variation is greater than 0.01 then the PDFs are considered not marginally calibrated \citep{galpro-github}. 

We use the methods described above to validate the redshift estimates and PDFs produced by GALPRO, as presented in the following sections.

 \subsection{Calibration}

The broadband magnitudes and their associated errors are used as input variables for GALPRO. Once the Zhou et al. truth table has been converted to obtain these parameters it is then randomly split. One million galaxies are randomly sampled from the truth table, with 90\% of the data being used to train the RF algorithm, and 10\% being used for testing and validation. GALPRO takes 4 input arrays, $x_{\text{train}}$ and $x_{\text{test}}$, which contain the luptitudes and parameters of the training and testing samples, and $y_{\text{train}}$ and $y_{\text{test}}$, containing the corresponding spectroscopic redshifts (the target variables).

\begin{figure}
  \centering
  \begin{tabular}{@{}c@{}}  \includegraphics[width=0.6\linewidth]{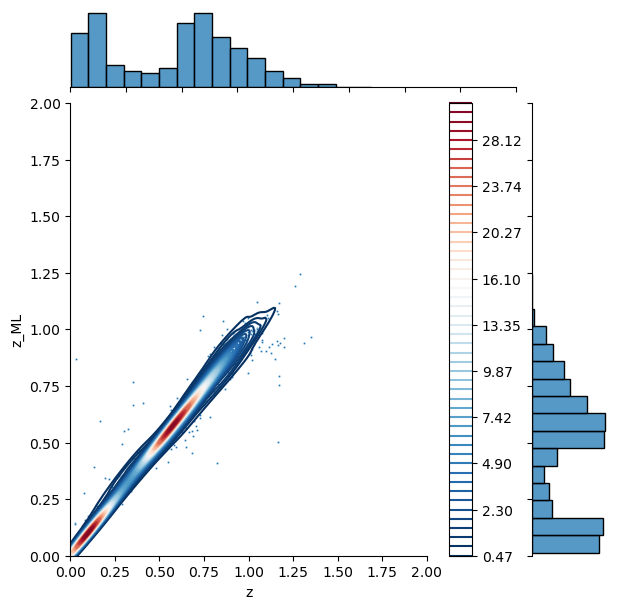} \\[\abovecaptionskip]
    \small (a) Photometric vs spectroscopic redshift plot. 
  \end{tabular}

  \vspace{\floatsep}

   \centering
  \begin{tabular}{@{}c@{}}
\includegraphics[width=0.65\linewidth]{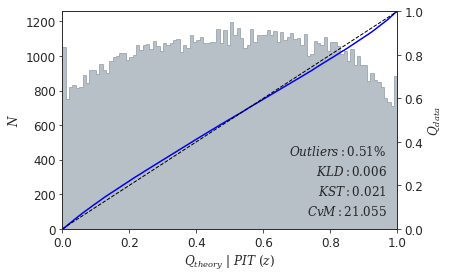} \\[\abovecaptionskip]
    \small (b) PIT plot, also showing the outlier fraction,\\ 
    KLD, KST and CvM test statistics.
  \end{tabular}

    \vspace{\floatsep}

    \centering
  \begin{tabular}{@{}c@{}}
\includegraphics[width=0.6\linewidth]{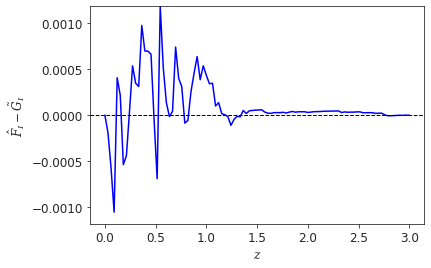} \\[\abovecaptionskip]
    \small (c) Marginal calibration plot.
  \end{tabular}
  
  \caption{GALPRO results obtained when trained and tested using the Zhou et al. calibration sample. In the upper panel the dots show the photometric redshift estimate (vertical axis) versus the true spectroscopic redshift (horizontal axis) for galaxies in the testing sample. Also plotted are coloured contours representing the joint likelihood of the photometric and spectroscopic redshifts, with the contour colour bar  displaying log likelihood values. Above and to the right of the plot are histograms of the redshift distributions for the training and testing samples respectively. The middle panel plots the PIT for the calibration sample and also shows the outlier fraction and the values of the KLD, KST and CvM test statistics. The lower panel shows the marginal calibration plot for the calibration sample.}\label{calib}
\end{figure}

The GALPRO analysis was run and GALPRO was initially calibrated using the Zhou sample to produce marginally and probabilistically calibrated redshift PDFs in the range $0 < z < 1.5$. The photometric versus spectroscopic redshift plot (upper panel) and the PIT (middle panel) and marginal calibration (lower panel) plots for the Zhou et al. sample are shown in Figure \ref{calib}. In the upper panel the dots show the photometric redshift estimate (vertical axis) versus the true spectroscopic redshift (horizontal axis) for galaxies in the testing sample\footnote{For greater clarity, in this plot, and in the equivalent plots in later figures, the dots are randomly downsampled to show only 10\% of the galaxies in the testing sample.}. Also plotted are coloured contours representing the joint likelihood of the photometric and spectroscopic redshifts, 
with the contour colour bar  displaying log likelihood values. Shown above and to the right of the plot are histograms of the redshift distributions for the training and testing samples respectively.

In Figure \ref{calib} the photometric versus spectroscopic redshift plot shows a strong correlation between the estimated photometric redshifts and the true redshift values, consistent with the estimates indeed being reliable. The value of $\sigma_{\text{NMAD}}$ characterises the deviation of the datapoints from an exact match between the spectroscopic and photometric redshift values. In this case the data give $\sigma_{\text{NMAD}} = 0.029$, which is relatively low, consistent with the strong correlation. Above a spectroscopic redshift value of around $z = 1.5$, however, GALPRO tends to underestimate the photometric redshift value as the training sample did not contain very many objects at these higher redshifts. This highlights the importance of representative training samples when using an RF algorithm for prediction.

The marginal calibration plot for the Zhou et al. sample can be seen in the lower panel of Figure \ref{calib}. There are negligible fluctuations about the zero line with a peak deviation of around 0.001, demonstrating that the PDFs generated using this testing sample are successfully marginally calibrated.

It was found that scattering the photometry also reduced the outlier fraction from 1.15\% to 0.51\%, which is very reasonable as the PIT (middle panel) is seen to be fairly uniform. On the other hand, the CvM test gave a value of 21.055, which is reflective of the fact that the PIT is slightly convex. 

The magnitudes were scattered in each filter and scattered colours were computed. Each galaxy in the training dataset was scattered five times and in the testing dataset scattered once for consistency. This produces marginally and probabilistically calibrated redshift PDFs, making the dataset a good choice for the trusted truth table. Any future reference to the truth dataset in this work refers to the scattered Zhou et al. survey.

The full calibration process produces redshift point estimates, PDFs, the PIT, marginal calibration and spectroscopic versus photometric redshift plots. GALPRO can also produce Kendall calibration plots -- see \citet{galpro-github} for further details. However these plots are used to assess the calibration of the joint posterior distributions, which are not particularly relevant to the motivation of this work. 

The production and storage of the redshift PDFs and point estimates only takes a short time, around 20 minutes in total, while the marginal and probabilistic calibration takes up the majority of the 6-hour period. This long calibration time may seem impractical, but it is worth noting that this calibration only needs to be run once using a representative subsample of the testing data to ensure accurate results. After this, GALPRO can be applied to the entire testing sample and run in a way which only produces the redshift PDFs, thus reducing the run-time significantly.

\section{Applying GALPRO to New Survey Data}\label{section_3}

\subsection{Overlap Tests} \label{overlap}

To investigate how statistically equivalent the training and testing samples must be in order for the RF generated by GALPRO to reliably estimate photometric redshifts in the training sample, the Zhou et al. sample was split into two different artificially-generated subsamples: one used for training and the other for testing.  This approach aimed to simulate the situation in which there are two distinctly different photometric surveys used for training and testing respectively. The extent of the overlap between the testing and training surveys -- in terms of the ranges and distributions of their observed photometry -- could then be varied, from the most extreme case with no overlap in range, to that of a complete overlap. 

In what follows, our analysis was implemented by training using `Survey 1', which was restricted to cover a specified range in the $r$-band magnitude distribution, and testing using `Survey 2' which was also restricted to a certain $r$-band magnitude range. These two surveys were generated with 90\%, 80\% and 70\% statistical overlap between the two $r$-band magnitude distributions, and the performance of GALPRO was evaluated in each case. For the 80\% overlap case, for example, the Zhou et al. galaxies were randomly assigned to Survey 1 and Survey 2 via the following steps:

\begin{itemize}
\item Identify the 80\% overlap region as the range of magnitudes that lies between the 10\% ($r_{10}$) and 90\% ($r_{90}$) percentiles in the CDF of the $r$-band magnitude distribution.
\item Construct Survey 1 by first randomly sampling $X_1$ galaxies from those which lie within the 80\% overlap region, between $r_{10}$ and $r_{90}$, and then sampling the remaining $Y_1$ galaxies from those for which $r < r_{10}$. 
\item Construct Survey 2 by first randomly sampling $X_2$ galaxies from within the 80\% overlap region, between $r_{10}$ and $r_{90}$, and the remaining $Y_2$ galaxies from those for which $r > r_{90}$.
\end{itemize}

In other words, Survey 1 was generated from all of the galaxies that lie within or below this 80\% overlap range in $r$-band magnitude, while Survey 2 was generated from all of the galaxies that lie within or above the overlap range. A similar approach can be used to construct surveys with 90\% and 70\% statistical overlap -- i.e.~in each case assuming that the statistical overlap between the two surveys comprises the central part of the CDF of magnitude values.  Clearly, many other ways to define the region(s) of statistical overlap are also possible but we do not consider them further here. 


\subsubsection{Basic Validations}
\label{sancheck}

As a sanity check to assess the validity of the above general approach, two limiting test cases were initially analysed.  Appendix \ref{sane} presents summary plots of the results for these two cases.

The first case involves training and testing the algorithm with subsamples that are completely statistically equivalent -- i.e.~with 100\% overlap and identical CDFs. This was done by randomly dividing the Zhou et al. dataset into two subsets -- referred to as `Survey 1' and `Survey 2' as before -- but with each drawn from exactly the same parent distribution. Consequently these two surveys were generated from identical ranges and distributions for their photometry in every band. Without loss of generality, Survey 1 was used as the training sample and Survey 2 was used as the testing sample. Entirely as expected, when the GALPRO analysis was run using these completely overlapping surveys, accurate redshift estimates with a uniform PIT were obtained in all cases -- consistent with the training and testing samples indeed being statistically equivalent and displaying no underlying difference in their features.

The second validation carried out explored how the RF algorithm performs in the extreme case where Survey 1 and Survey 2 have {\em no\/} overlapping range and therefore have completely disjoint statistical distributions. This check was carried out by splitting the Zhou et al. truth dataset in half, by identifying the median of the $r$-band magnitude range, and then taking Survey 1 to comprise all of the galaxies with $r \leq r_{\rm median}$ while Survey 2 comprised all of the galaxies with $r \geq r_{\rm median}$. The same GALPRO analysis was then run and, unsurprisingly, yielded wholly inaccurate results -- with the PIT plot indicating that a large degree of bias had been introduced.

In summary, then, the spectroscopic versus photometric redshift, PIT and marginal calibration plots for both limiting sanity check cases behaved exactly as expected: when the randomly generated `surveys’ used for training and testing were drawn from identical statistical distributions they gave well-calibrated photometric redshift PDFs. On the other hand, when the surveys used for training and testing were drawn from completely disjoint distributions, the estimated photometric redshifts were mainfestly unsuitable and were essentially invalid.

With the above basic validations successfully carried out, we next used the Zhou et al. sample to generate artificially Survey 1 and Survey 2 data, for training and testing purposes respectively, with a 90\%, 80\% and 70\% statistical overlap in their $r$-band photometry. 

\subsubsection{Results: 90\% Overlap}

\begin{figure}
  \centering
  \begin{tabular}{@{}c@{}}
\includegraphics[width=.6\linewidth]{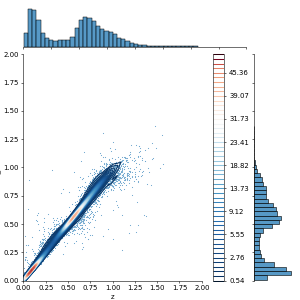} \\[\abovecaptionskip]
    \small (a) Photometric versus spectroscopic redshift plot.
  \end{tabular}

  \vspace{\floatsep}
    \centering
  \begin{tabular}{@{}c@{}}
\includegraphics[width=.65\linewidth]{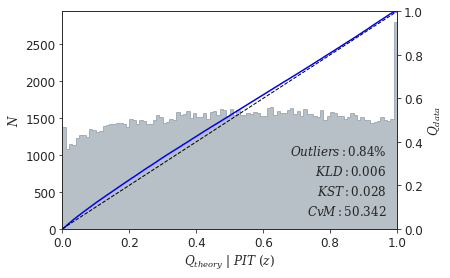} \\[\abovecaptionskip]
    \small (b) PIT plot, also showing the outlier fraction,\\ 
    KLD, KST and CvM test statistics.
  \end{tabular}

    \vspace{\floatsep}

     \centering
  \begin{tabular}{@{}c@{}}
\includegraphics[width=.6\linewidth]{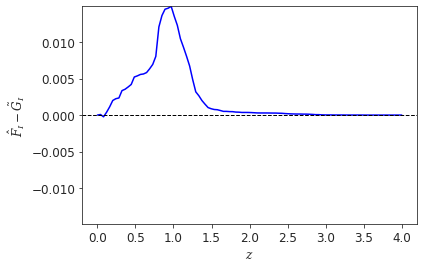} \\[\abovecaptionskip]
    \small (c) Marginal calibration plot.
  \end{tabular}
  
  \caption{GALPRO results when trained using the Zhou et al. sample and adopting a 90\% overlap between the training and testing datasets. In the upper panel the dots show the photometric redshift estimate (vertical axis) versus the true spectroscopic redshift (horizontal axis) for galaxies in the testing sample. Also plotted are coloured contours representing the joint likelihood of the photometric and spectroscopic redshifts, with the contour colour bar  displaying log likelihood values. Above and to the right of the plot are histograms of the redshift distributions for the training and testing samples respectively. The middle panel plots the PIT for the calibration sample and also shows the outlier fraction and the values of the KLD, KST and CvM test statistics. The lower panel shows the marginal calibration plot for the calibration sample.}\label{90}
\end{figure}

First, GALPRO was trained using Survey 1 and tested using Survey 2, with a 90\% overlap between the $r$-band photometry distribution of the two surveys and otherwise using the same settings as in all of the previous tests. The photometric versus spectroscopic redshift scatter and contour plot (upper panel), PIT plot (middle panel) and marginal calibration plot (lower panel) can be seen for this case in Figure \ref{90}. The PIT produced by this test was generally uniform; however, it does display a small downward dip at lower values. The outlier fraction, KLD and KST tests all gave reasonable values, although the CvM test gave a value of 50.342 indicating that some bias was clearly present in the results. The marginal calibration plot lies entirely above the zero line instead of randomly oscillating about zero, although its peak value of around 0.015 is still consistent with a successful marginal calibration. The photometric versus spectroscopic redshift data again gave a value of $\sigma_{\rm NMAD} = 0.029$, which is again reasonable. However, the plot itself also shows that some small amount of bias has been introduced. 

From the upper panel of Figure \ref{90} it is clear that there is a strong correlation between the photometric and spectroscopic redshifts over the range $0.5 < z < 1$. However at larger $z$ values, it is also clear that GALPRO underestimates the photometric redshifts. Due to the $r$-band magnitude cut-off, the training sample contains little to no galaxies with a redshift $z > 0.9$, whereas the testing sample reaches out to a redshift of 1.5. The underestimation of photometric redshifts at these higher values of the true redshift demonstrates that GALPRO appears only able to learn the mapping between the fluxes and redshifts over the range present in the training sample and is less able to extrapolate the mapping to higher/lower redshift values. Indeed it is not only the range, but also the shape, of the redshift distributions of the testing and training samples which seems to affect the results. For $z < 0.5$, the redshift values produced by GALPRO are slightly overestimated. The testing sample does contain galaxies at lower redshifts; however, the shape of the two redshift distributions for $z < 0.5$ is different, which introduces a bias in the estimated photometric redshifts. Our analysis was then run again, this time with Survey 1 now used for testing and Survey 2 for training. This gave very similar results, with equally well-calibrated redshift estimates and PDFs, providing further confidence in the robustness of our results. 

Overall, then, when there is a 90\% overlap in the distribution of $r$-band photometry between the testing and training samples, our results find that the redshift estimates and PDFs obtained for the training sample are sufficiently well-calibrated and valid to be useful. 

\subsubsection{Results: 80\% Overlap}

\begin{figure}
  \centering
  \begin{tabular}{@{}c@{}}
\includegraphics[width=.6\linewidth]{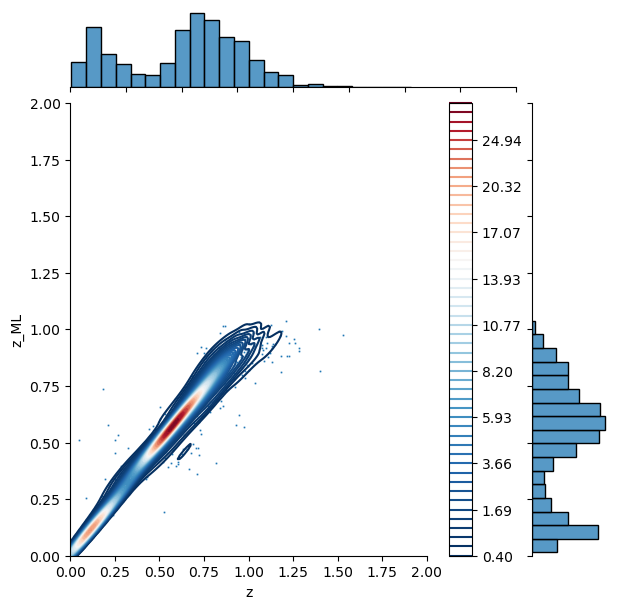} \\[\abovecaptionskip]
    \small (a) Photometric versus spectroscopic redshift plot.
  \end{tabular}

  \vspace{\floatsep}
    \centering
  \begin{tabular}{@{}c@{}}
    \includegraphics[width=.65\linewidth]{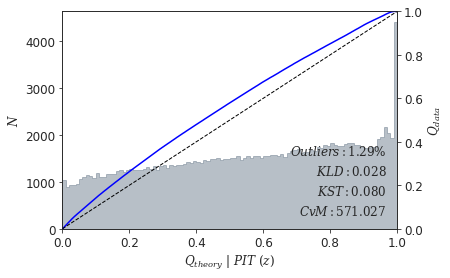} \\[\abovecaptionskip]
    \small (b) PIT plot, also showing the outlier fraction,\\ 
    KLD, KST and CvM test statistics.
  \end{tabular}

    \vspace{\floatsep}

   \centering
  \begin{tabular}{@{}c@{}}
    \includegraphics[width=.6\linewidth]{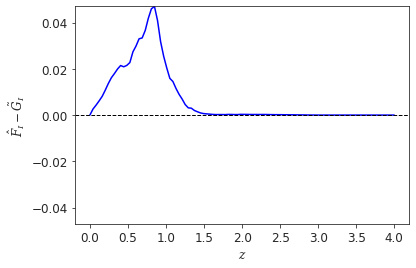} \\[\abovecaptionskip]
    \small (c) Marginal calibration plot
  \end{tabular}
  
  \caption{GALPRO results when trained using the Zhou et al. sample and adopting an 80\% overlap between the training and testing datasets. In the upper panel the dots show the photometric redshift estimate (vertical axis) versus the true spectroscopic redshift (horizontal axis) for galaxies in the testing sample. Also plotted are coloured contours representing the joint likelihood of the photometric and spectroscopic redshifts, with the contour colour bar  displaying log likelihood values. Above and to the right of the plot are histograms of the redshift distributions for the training and testing samples respectively. The middle panel plots the PIT for the calibration sample and also shows the outlier fraction and the values of the KLD, KST and CvM test statistics. The lower panel shows the marginal calibration plot for the calibration sample.}\label{80}
\end{figure}

Next, the training and testing samples were split so that Survey 1 and Survey 2 had an 80\% overlap region in their $r$-band photometry. GALPRO was then trained using Survey 1 and tested using Survey 2, with the same hyperparameters as used in all of the previous tests. The photometric versus spectroscopic redshift scatter and contour plot, PIT and marginal calibration plots can be seen for this case in Figure \ref{80}. 

The PIT plot, in the middle panel of Figure \ref{80}, shows that a significant bias has been introduced to the PDFs, and moreover the plot has a noticeably steeper gradient -- indicating a stronger departure from uniform behaviour -- than for the PIT plot in the case of 90\% overlap. The outlier fraction has also increased compared with the 90\% overlap case and the CvM test has now increased by about an order of magnitude to give a value of 571.027, indicating very clearly that probabilistic calibration has not been successful. The marginal calibration plot is again systematically positive and peaks at a value of more than 0.04, indicating that the PDFs are not marginally calibrated. This peak value is more than three times larger than the peak found in the marginal calibration plot for the 90\% overlap test, indicating that both the marginal and probabilistic calibrations have significantly worsened as the overlap percentage between the testing and training samples decreases from 90\% to 80\%. The photometric versus spectroscopic redshift data, shown in the upper panel, give a value of $\sigma_{\rm NMAD} = 0.036$, indicating that the scatter has also worsened as the overlap has decreased. Again, out to large values of the spectroscopic redshift ($z > 1$), the photometric redshift estimates are seen to become very inaccurate, and GALPRO fails to predict any estimated redshifts larger than about $z \sim 1$. This again is consistent with the behaviour that GALPRO is unable to successfully extrapolate the learnt mapping to redshifts outside the range that are contained in the training sample. 

Even where the training and testing samples have an overlapping range of redshifts -- i.e.~for $0.4 < z_{\rm spec} < 0.6$ -- the photometric redshifts estimated by GALPRO were found to be noticeably less accurate with an 80\% overlap of the $r$-band magnitude distributions for the two surveys, compared with our results for the previous 90\% overlap test.  The reason for this would appear to derive from the shape of the redshift distributions in each survey. As the two surveys have redshift distributions of different shape within their overlapping range, it appears that the mapping between the fluxes and redshifts learned for Survey 1 cannot be successfully applied to Survey 2. This is disappointing, as it seems to indicate that not only does the range of the magnitude distributions have to be very similar for each survey, but also the shape of their redshift distributions within the region of overlap.

Seeking to apply GALPRO to estimate photometric redshifts from a different survey for which there is an 80\% overlap in $r$-band magnitudes does not seem an unreasonable scenario, and would not be dissimilar to the sort of overlap we might expect to see in practice -- a point we return to in Section~\ref{summary_conc}. Nevertheless, the GALPRO RF does not perform to an acceptable standard in this scenario.

As before, our analysis was also run with Survey 1 used for testing and Survey 2 for training. Again we found that our results were consistent.

\subsubsection{Results: 70\% Overlap}

Finally, the training and testing samples were split so that they now had only a 70\% overlap in their $r$-band magnitude distribution. Survey 1 was again used for training and Survey 2 for testing. 

\begin{figure}
  \centering
  \begin{tabular}{@{}c@{}}   \includegraphics[width=0.6\columnwidth]{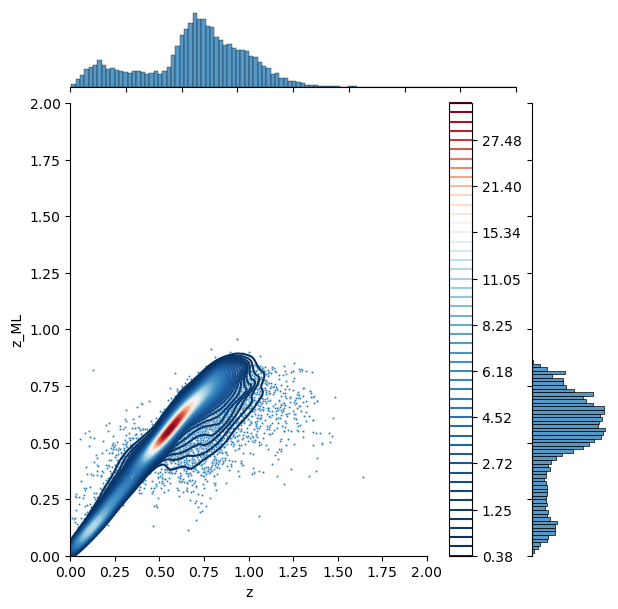} \\[\abovecaptionskip]
    \small (a) Photometric versus spectroscopic redshift plot.
  \end{tabular}

  \vspace{\floatsep}
  
  \centering
  \begin{tabular}{@{}c@{}}
    \includegraphics[width=.65\linewidth]{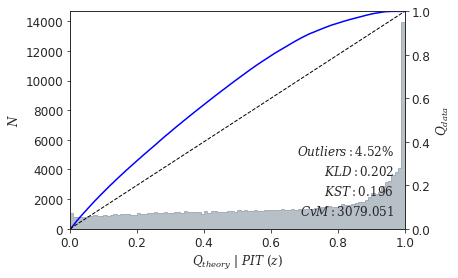} \\[\abovecaptionskip]
    \small (b) PIT plot, also showing the outlier fraction,\\ 
    KLD, KST and CvM test statistics.
  \end{tabular}

    \vspace{\floatsep}

   \centering
  \begin{tabular}{@{}c@{}}
    \includegraphics[width=.6\linewidth]{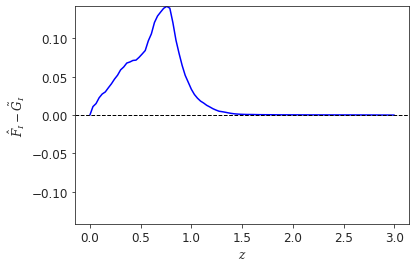} \\[\abovecaptionskip]
    \small (c) Marginal calibration plot
  \end{tabular}
  
  \caption{GALPRO results when trained using the Zhou et al. sample and adopting a 70\% overlap between the training and testing datasets. In the upper panel the dots show the photometric redshift estimate (vertical axis) versus the true spectroscopic redshift (horizontal axis) for galaxies in the testing sample. Also plotted are coloured contours representing the joint likelihood of the photometric and spectroscopic redshifts, with the contour colour bar  displaying log likelihood values. Above and to the right of the plot are histograms of the redshift distributions for the training and testing samples respectively. The middle panel plots the PIT for the calibration sample and also shows the outlier fraction and the values of the KLD, KST and CvM test statistics. The lower panel shows the marginal calibration plot for the calibration sample.}\label{70}
\end{figure}

As expected, this test produced highly inaccurate photometric redshift estimates, which are presented in Figure \ref{70}, continuing the trend that was apparently in the previous results for 90\% and 80\% overlap respectively. In the middle panel of Figure \ref{70} the PIT plot shows that a catastrophic degree of bias has been introduced, with a very steep gradient and a CvM test statistic value of 3079.051. In fact, it appears that with each reduction of 10\% in survey overlap, the CvM value increases by about a factor of ten. The marginal calibration plot, shown in the lower panel, now peaks at a value of about 0.15, which again is much larger than for the 80\% overlap test. Moreover, the plot again is systematically positive and clearly indicates that the marginal calibration is wholly unsuccesful. Finally, the spectroscopic versus photometric redshift plot shown in the upper panel is also significantly poorer, with much lower correlation and the data giving a value of $\sigma_{\rm NMAD} \sim 0.05$. Furthermore, GALPRO fails to predict any photometric redshift estimates greater than $z \sim  0.85$, which corresponds with the cut-off in spectroscopic redshift of the training sample. This again reinforces the pattern that GALPRO may only predict photometric redshift values that lie within the limits of the training sample used.

Hence, the decrease in overlap by a further 10\% has further degraded the accuracy of the redshift estimation and calibration by a significant amount and clearly demonstrates that the degree of overlap between the training and testing samples can substantially affect the performance of the RF. Although an overlap of only 70\% in $r$-band magnitude might seem to be a realistic expectation when applying GALPRO to a new survey, it is clear that this degree of mismatch is already enough to cause a complete failure in the RF algorithm, leading to inaccurate, biased and unreliable results.

As before, our analysis was run again, with Survey 1 this time used for testing and Survey 2 for training, and again a consistent pattern of results was obtained. 

Overall, then, we find that the training and testing datasets given to GALPRO should overlap by at least 90\% in their $r$-band magnitude distributions in order to produce marginally and probabilistically calibrated redshift PDFs that are acceptable.

\subsection{Applying GALPRO to a Completely Different Survey}\label{pan_section}

We next investigate whether GALPRO can be reliably used to generate photometric redshifts when applied to a completely new galaxy catalogue for which no spectroscopic calibrating data may exist. 

Here, we use the Panoramic Survey Telescope and Rapid Response System (PanSTARRS) survey to play the role of our `new’ survey, which hypothetically has no spectroscopic data. In fact, the PanSTARRS survey has been cross-matched to provide a large sample of spectroscopic redshifts which can be used for validation \citep{89}. Consequently we can use these PanSTARRS spectroscopic redshifts to assess the reliability of the photometric redshift estimates that are generated from the PanSTARRS photometric data using GALPRO trained with the Zhou et al. sample. 

The PanSTARRS telescope is located in Hawaii and accurately measures photometry of known objects by surveying for variable objects using telescopes, cameras and a large computer system \citep{chambers2019panstarrs1}. The main objective of the PanSTARRS survey is to observe objects that are near to the Earth and which may pose the threat of impact. As a result, the survey is relatively shallow. A sample of around 2 million galaxies included in the PanSTARRS survey have cross-matched spectroscopic redshifts, and these redshifts can be used to assess the performance of the applied RF.



It is useful to compare the spectroscopic redshift distributions for the two surveys. The CDFs of the PanSTARRS and DESI datasets have a KST statistic between them of 0.0527, which is consistent with there being no statistically significant difference between the two distributions. In fact, it is interesting to note that the Zhou et al. and PanSTARRS datasets have spectroscopic redshift distributions that are {\em more\/} similar (as characterised by their KST statistic) than the randomly-generated redshift distributions of Survey 1 and Survey 2 in the 90\% overlap case considered in Section \ref{90}.  In that case the two spectroscopic redshift distributions sampled had a KST statistic of 0.198. 


Using PanSTARRS to represent our `new' photometric survey presents a potentially challenging test for GALPRO trained with the Zhou et al. sample, since the two surveys use different photometric systems. Firstly, the $g$, $r$ and $z$ bands are defined differently between the two surveys and the corrections summarised in \cite{102} must be applied to the data in these photometric bands. Secondly, the $W$ magnitude systems of the two datasets are also different as the PanSTARRS survey defines the $W$ bands using the \textit{vega} system, whereas the DESI Legacy survey uses the $ab$ system. However, the PanSTARRS $W$ bands can in principle be converted to the $ab$ system using the following relation\footnote{See {https://wise2.ipac.caltech.edu} for further details.}:
\begin{equation}
   m_{ab} = m_{\rm vega} + \Delta m ,
\end{equation}
with
\begin{equation}
\Delta m  =  
\begin{cases}
    2.699 & \text{for the $W1$ band} \\
    3.339 & \text{for the $W2$ band}, \\
\end{cases}
\end{equation}
where $m_{ab}$ is the $ab$ magnitude, and $m_{\rm vega}$ is the Vega magnitude.

GALPRO was trained using a random sample of 450,000 galaxies from the scattered truth table, with the hyperparameters set to those previously described in Section \ref{hyper}. A random sample of 250,000 galaxies from the PanSTARRS survey was then used as the testing dataset. The training and testing datasets consist of the $r$-$z$, $g$−$r$, $z$-$W1$ and $W1$-$W2$ bands (corrected as appropriate) and their errors, drawn from the respective surveys, with the spectroscopic redshift data from the PanSTARRS survey used for validation. 

\begin{figure}
  \centering
  \begin{tabular}{@{}c@{}}
\includegraphics[width=0.6\linewidth]{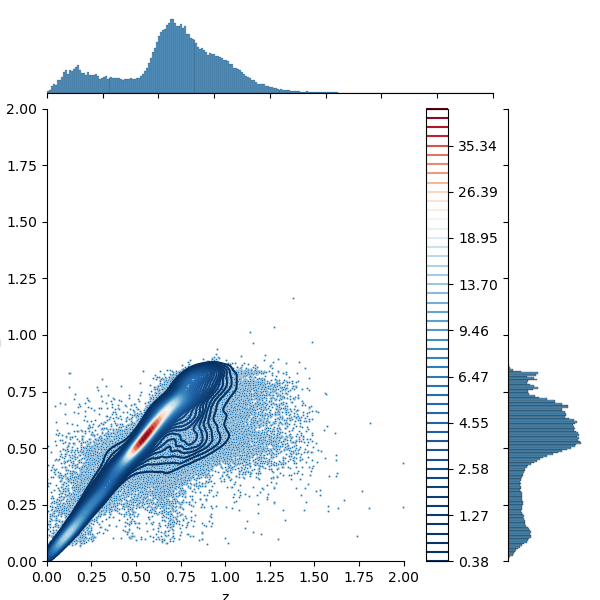} \\[\abovecaptionskip]
    \small (a) Photometric versus spectroscopic redshift plot.
  \end{tabular}

  \vspace{\floatsep}

\centering
  \begin{tabular}{@{}c@{}}
\includegraphics[width=0.65\linewidth]{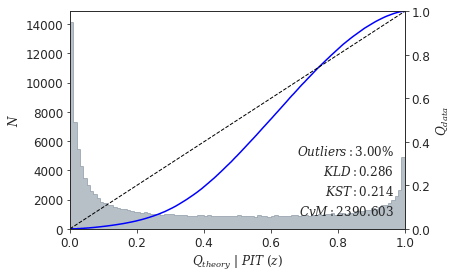} \\[\abovecaptionskip]
    \small (b) PIT plot, also showing the outlier fraction,\\ 
    KLD, KST and CvM test statistics.
  \end{tabular}

    \vspace{\floatsep}

   \centering
  \begin{tabular}{@{}c@{}}
    \includegraphics[width=0.6\linewidth]{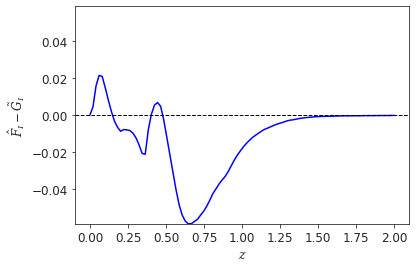} \\[\abovecaptionskip]
    \small (c) Marginal calibration plot
  \end{tabular}
  
  \caption{GALPRO results when trained using the Zhou et al. sample and tested using the PanSTARRS survey. In the upper panel the dots show the photometric redshift estimate (vertical axis) versus the true spectroscopic redshift (horizontal axis) for galaxies in the testing sample. Also plotted are coloured contours representing the joint likelihood of the photometric and spectroscopic redshifts, with the contour colour bar  displaying log likelihood values. Above and to the right of the plot are histograms of the redshift distributions for the training and testing samples respectively. The middle panel plots the PIT for the calibration sample and also shows the outlier fraction and the values of the KLD, KST and CvM test statistics. The lower panel shows the marginal calibration plot for the calibration sample. }\label{final}
\end{figure}

The results of this analysis are summarised in Figure \ref{final}.  The upper panel shows a plot of the photometric redshift estimates versus the spectroscopic redshifts for the randomly selected PanSTARRS survey galaxies in the testing sample. As in previous figures, also plotted are coloured contours representing the joint likelihood of the photometric and spectroscopic redshifts, 
with the contour colour bar  displaying log likelihood values, and shown above and to the right of the plot are histograms of the redshift distributions for the training and testing samples respectively.  The middle panel shows the PIT for the PanSTARRS survey, together with the outlier fraction, KLD, KST and CvM statistic values, and finally the lower panel shows the marginal calibration plot for the PanSTARRS survey galaxies. 

These results clearly show that the GALPRO RF, trained on the Zhou et al. sample, has {\em not\/} been successfully transferred to the PanSTARRS testing sample. The PIT is extremely concave, indicating that the PDFs are systematically too narrow by a significant amount. The CvM test yields a value of 2390.603, consistent with this concavity. The marginal calibration does oscillate about the zero line, but predominantly below that line and in a smooth manner -- indicative of a systematic negative bias. It peaks at a negative value of around 0.06, indicating that the marginal calibration has not been successful. In the upper panel, while the plot of photometric versus spectroscopic redshifts does generally follow the diagonal, the scatter about the diagonal is significant and yields a value of $\sigma_{\rm NMAD} = 0.060$. There is no clear cut off in the scatter plot, which is as would be expected since the redshift range of the training sample encapsulates the redshift range of the PanSTARRS sample. However, the large value of $\sigma_{\rm NMAD}$ obtained indicates that the photometric redshift estimates are inaccurate, and (as already noted) the PIT and marginal calibration graphs show that the PDFs are not probabilistically or marginally calibrated.
The overlap tests described in Section \ref{overlap} showed that a discrepancy in the range of the photometric bands between the training and testing samples will tend to introduce a bias in the PDFs, which appears as a gradient in the PIT plot. In Figure \ref{final}, however, we see that the PIT does not have a uniform gradient but is concave in shape, indicating that the poor performance of GALPRO is not related to the overlap (or lack thereof) of the photometric bands. Instead, the PDFs inferred for the PanSTARRS galaxies are too narrow by a significant amount.  We discuss these results further in Section \ref{summary_conc}.

\section{Discussion and Conclusions}
\label{summary_conc}

This work has served as an investigation of an RF-based approach, GALPRO, to estimate photometric redshifts from galaxy survey data, and to evaluate whether GALPRO, trained using one galaxy survey, can be reliably applied to estimate photometric redshifts for a new survey comprising only photometric data, i.e.~in the absence of spectroscopic redshifts in the new survey that can be directly used for training purposes.

GALPRO was initially calibrated using a dataset compiled from the DESI legacy survey and our analysis confirmed that GALPRO produced accurate photometric redshift PDFs, demonstrating that the software is useful when applied to a galaxy catalogue that contains an incomplete sample of spectroscopic redshifts for the purpose of augmenting and supplementing those spectroscopic redshifts with estimated photometric redshifts. 

Analysis was then carried out to investigate the case where GALPRO was applied to a different galaxy survey, in order to determine how the overlap in the ranges of the data in the two surveys affects the performance of the algorithm. It was found that, even where the two surveys overlapped by 90\% in their $r$-band magnitude range, the results obtained with GALPRO displayed a small but noticeable amount of bias when compared with the results of the initial calibration tests. The tests also indicated that not only the range of of the data in the given photometric bands, but also the shape of the redshift distributions should be very similar for the RF to be applicable to the second survey. This demonstrates that GALPRO may not be suitable for the estimation of photometric redshifts when applied to an entirely new survey of unknown (but likely greater) depth and range, as GALPRO cannot extrapolate the mapping it has learnt between fluxes and redshift outside of the redshift range of the training sample.

To further assess how well GALPRO works for new surveys for which we have {\em only\/} photometric data available, the PanSTARRS galaxy catalogue was then chosen as the testing sample, i.e.~acting as a `new’ survey for which we only have photometry available. Since PanSTARRS uses a different magnitude system for its photometric bands, the PanSTARRS magnitudes first had to be converted to the system used by the DESI legacy survey.

Despite the PanSTARRS and Zhou et al. samples having consistent and strongly overlapping ranges of both their spectroscopic redshifts and their (suitably corrected) photometric data, when the Zhou et al. sample was used for training and the PanSTARRS sample for testing the photometric redshifts derived for the testing sample were {\em not\/} successfully calibrated. Our results indicate that the mapping learnt between the colours and redshifts of the Zhou et al. sample is therefore {\em not\/} transferable to the PanSTARRS sample.

The origin of the poor performance of GALPRO when applied to PanSTARRS is not entirely clear but is most likely a result of the different photometric systems used by the two surveys. The use of the luptitude system in the Zhou et al. sample (and in particular its adoption of an inverse hyperbolic function to relate photometric fluxes and magnitudes) is not in itself responsible for a  mismatch between the training and testing data, since the inverse hyperbolic transformation is almost identical to the standard logarithmic relationship for high signal-to-noise observations. However, the transformation relations used to convert the magnitudes between the two systems (as briefly summarised in Section \ref{pan_section}) will introduce an intrinsic scatter to the PanSTARRS data that adds a significant`noise floor' to the estimated photometric redshifts. It would appear that this noise floor is sufficient to significantly degrade the performance of the mapping between the fluxes and redshifts learned from the training data when it is applied to the new survey. 

Insofar as they are applicable to other survey datasets and to other algorithms and approaches to determining photometric redshifts, our findings may have implications for other recent ML-based analyses in the literature that derive photometric redshifts through combining photometry from multiple surveys.  In particular, our results in Section \ref{overlap} -- concerning the transferablity of photometric redshift estimation methods when applied to different surveys with progressively smaller overlap in their photometry distributions -- could be of relevance to these analyses.

We note that \cite{2020MNRAS.493.2059B} applied the photometric redshift fitting code Le-Phare \citep{1999MNRAS.310..540A,2006A&A...457..841I} to derive photometric redshifts for galaxies observed in the COSMOS and XMM-Newton Large Scale Structure fields, using deep near-infrared photometry carried out for the UltraVISTA fourth data release \citep{2012A&A...544A.156M} and the VIDEO survey \citep{2013MNRAS.428.1281J} respectively. However, our results on the limited `transferability' from training to testing data of photometric redshifts estimated using ML-based methods will not apply to template-based techniques such as LePhare.

In \cite{2022MNRAS.513.3719H} the authors explicitly adopt a `hybrid' approach, similar to that considered in this work, to investigate the performance of photometric redshift estimates, derived using the GPz supervised ML-based method \citep{RasmussenW06}, trained using spectroscopic redshift data obtained in one survey field and then applied to another.  The results presented in \cite{2022MNRAS.513.3719H} appear to suggest that the mapping between the fluxes and redshifts derived in the training data {\em is\/} transferable to another survey -- albeit with some bias introduced, and with a value of $\sigma_{\text{NMAD}}$ and an outlier fraction which are somewhat {\em larger\/} than the values that we obtained when GALPRO was transferred from the Zhou et al. sample to PanSTARRS.  It would be interesting to compute PIT and marginal calibration plots for the testing data investigated in \cite{2022MNRAS.513.3719H} and to compare these with the corresponding plots shown in Figure \ref{final} of this work.

Whether or not there are any residual issues around the transferability of the machine learning-based photometric redshifts considered in \cite{2022MNRAS.513.3719H}, the hierarchical Bayesian methodology presented in that paper -- which seeks to determine an optimal `consensus' posterior probability distribution for the photometric redshift for each galaxy, based on combining information from multiple distinct estimates - would seem to be a useful and powerful approach to adopt in situations where the applicability of a photometric redshift estimator to a new survey might be a concern. Such approaches will be very important in the future when considering how to combine optimally photometric data from e.g. Rubin and Euclid surveys.

Another relevant recent study that explores the transferability of photometric redshifts to new surveys is 
\cite{ml4}. The authors use the TensorFlow\footnote{https://www.tensorflow.org} deep learning (DL) algorithm to investigate the accuracy of photometric redshifts, trained using a sample of about 70,000 quasi-stellar objects (QSOs) from the Sloan Digital Sky Survey, when applied to other radio-selected datasets. They conclude that their DL algorithm generally performs well at predicting photometric redshifts for the QSOs in the other datasets, across a wide range of radio fluxes. Unlike the PanSTARRS case considered in this work, however, the magnitudes used as inputs to the training and testing data considered were defined on consistent photometric systems.  Moreover, the authors also note that requiring the availability of consistent $u$, $g$, $r$, $i$, $z$, $W1$, $W2$, $NUV$ and $FUV$ magnitudes for both testing and training samples meant that reliable photometric redshifts could not be obtained for a significant fraction of the QSO sources in the radio-selected surveys.  Their findings would, therefore, appear to be consistent with our conclusion that the poor performance of GALPRO applied to PanSTARRS is most likely a result of the different photometric systems used by the two surveys. 

In summary, our results suggest that, in general, RF-based methods such as GALPRO may not be suitable for generating estimates of photometric redshifts and their PDFs using only photometric data from a different galaxy survey. Methods such as GALPRO may be useful in the case where a catalogue is nearly complete, but is missing some spectroscopic redshifts, since GALPRO can in this case be trained and tested using the same survey. However, the algorithm appears unable to extrapolate the learnt mapping between redshifts and colours to the new survey or to redshift values beyond the range over which it has been trained.

Our results should, therefore, serve as a cautionary note on the limits of the applicability and adaptability of RF-based algorithms for generating photometric redshifts -- particularly where they are applied to new surveys that use even slightly different photometric systems. Future work may include a deeper investigation into the calibration and application of RF-based methods between different galaxy catalogues, and an exploration of how software such as GALPRO might be extended to perform more reliably in these cases. 

\section*{Acknowledgments}
\vspace{-0.5em}
The authors give special thanks to Viv Kendon for her generous support and helpful comments. The authors also thank the anonymous referees whose comments have significantly improved the paper. M. H. is supported by the Science and Technology Facilities Council (Ref. ST/L000946/1). 

\section*{Data Availability}

The GALPRO software used in this paper is available from GitHub (https://github.com/smucesh/galpro). The datasets generated in the current study are available from the corresponding authors on reasonable request.



\bibliographystyle{mnras}
\bibliography{example} 

\begin{thebibliography}{}
\makeatletter
\relax
\def\mn@urlcharsother{\let\do\@makeother \do\$\do\&\do\#\do\^\do\_\do\%\do\~}
\def\mn@doi{\begingroup\mn@urlcharsother \@ifnextchar [ {\mn@doi@} {\mn@doi@[]}}
\def\mn@doi@[#1]#2{\def\@tempa{#1}\ifx\@tempa\@empty \href {http://dx.doi.org/#2} {doi:#2}\else \href {http://dx.doi.org/#2} {#1}\fi \endgroup}
\def\mn@eprint#1#2{\mn@eprint@#1:#2::\@nil}
\def\mn@eprint@arXiv#1{\href {http://arxiv.org/abs/#1} {{\tt arXiv:#1}}}
\def\mn@eprint@dblp#1{\href {http://dblp.uni-trier.de/rec/bibtex/#1.xml} {dblp:#1}}
\def\mn@eprint@#1:#2:#3:#4\@nil{\def\@tempa {#1}\def\@tempb {#2}\def\@tempc {#3}\ifx \@tempc \@empty \let \@tempc \@tempb \let \@tempb \@tempa \fi \ifx \@tempb \@empty \def\@tempb {arXiv}\fi \@ifundefined {mn@eprint@\@tempb}{\@tempb:\@tempc}{\expandafter \expandafter \csname mn@eprint@\@tempb\endcsname \expandafter{\@tempc}}}

\bibitem[\protect\citeauthoryear{{Almosallam}, {Jarvis}  \& {Roberts}}{{Almosallam} et~al.}{2016}]{2016MNRAS.462..726A}
{Almosallam} I.~A.,  {Jarvis} M.~J.,   {Roberts} S.~J.,  2016, \mn@doi [\mnras] {10.1093/mnras/stw1618}, \href {https://ui.adsabs.harvard.edu/abs/2016MNRAS.462..726A} {462, 726}

\bibitem[\protect\citeauthoryear{{Arnouts}, {Cristiani}, {Moscardini}, {Matarrese}, {Lucchin}, {Fontana}  \& {Giallongo}}{{Arnouts} et~al.}{1999}]{1999MNRAS.310..540A}
{Arnouts} S.,  {Cristiani} S.,  {Moscardini} L.,  {Matarrese} S.,  {Lucchin} F.,  {Fontana} A.,   {Giallongo} E.,  1999, \mnras, 310, 540

\bibitem[\protect\citeauthoryear{Beck et~al.,}{Beck et~al.}{2017}]{10.1093/mnras/stx687}
Beck R.,  et~al., 2017, Monthly Notices of the Royal Astronomical Society, 468, 4323

\bibitem[\protect\citeauthoryear{Benitez}{Benitez}{2000}]{photoz1}
Benitez N.,  2000, APJ, 536, 571

\bibitem[\protect\citeauthoryear{Biau \& Scornet}{Biau \& Scornet}{2016}]{rftour}
Biau G.,  Scornet E.,  2016, TEST, 25, 197

\bibitem[\protect\citeauthoryear{{Blake, C., et al.}}{{Blake, C., et al.}}{2011}]{wigglez}
{Blake, C., et al.} 2011, MNRAS, 415, 2892

\bibitem[\protect\citeauthoryear{{Bowler}, {Jarvis}, {Dunlop}, {McLure}, {McLeod}, {Adams}, {Milvang-Jensen}  \& {McCracken}}{{Bowler} et~al.}{2020}]{2020MNRAS.493.2059B}
{Bowler} R.~A.~A.,  {Jarvis} M.~J.,  {Dunlop} J.~S.,  {McLure} R.~J.,  {McLeod} D.~J.,  {Adams} N.~J.,  {Milvang-Jensen} B.,   {McCracken} H.~J.,  2020, \mnras, 493, 2059

\bibitem[\protect\citeauthoryear{Bozinovski}{Bozinovski}{2020}]{Bozinovski2020ReminderOT}
Bozinovski S.,  2020, Informatica (Slovenia), 44

\bibitem[\protect\citeauthoryear{Brammer, van Dokkum  \& Coppi}{Brammer et~al.}{2008}]{Brammer_2008}
Brammer G.~B.,  van Dokkum P.~G.,   Coppi P.,  2008, \mn@doi [APJ] {10.1086/591786}, 686, 1503

\bibitem[\protect\citeauthoryear{{Brescia}, {Cavuoti}, {D'Abrusco}, {Longo}  \& {Mercurio}}{{Brescia} et~al.}{2013}]{2013ApJ...772..140B}
{Brescia} M.,  {Cavuoti} S.,  {D'Abrusco} R.,  {Longo} G.,   {Mercurio} A.,  2013, \mn@doi [\apj] {10.1088/0004-637X/772/2/140}, \href {https://ui.adsabs.harvard.edu/abs/2013ApJ...772..140B} {772, 140}

\bibitem[\protect\citeauthoryear{{Carliles, S., et al}}{{Carliles, S., et al}}{2010}]{rfforphotoz}
{Carliles, S., et al} 2010, Astrophys. J.T, 712

\bibitem[\protect\citeauthoryear{{Carliles}, {Budav{\'a}ri}, {Heinis}, {Priebe}  \& {Szalay}}{{Carliles} et~al.}{2010}]{2010ApJ...712..511C}
{Carliles} S.,  {Budav{\'a}ri} T.,  {Heinis} S.,  {Priebe} C.,   {Szalay} A.~S.,  2010, \mn@doi [\apj] {10.1088/0004-637X/712/1/511}, \href {https://ui.adsabs.harvard.edu/abs/2010ApJ...712..511C} {712, 511}

\bibitem[\protect\citeauthoryear{{Carrasco Kind} \& {Brunner}}{{Carrasco Kind} \& {Brunner}}{2013}]{2013MNRAS.432.1483C}
{Carrasco Kind} M.,  {Brunner} R.~J.,  2013, \mn@doi [\mnras] {10.1093/mnras/stt574}, \href {https://ui.adsabs.harvard.edu/abs/2013MNRAS.432.1483C} {432, 1483}

\bibitem[\protect\citeauthoryear{{Cavuoti}, {Brescia}, {De Stefano}  \& {Longo}}{{Cavuoti} et~al.}{2015}]{ml6}
{Cavuoti} S.,  {Brescia} M.,  {De Stefano} V.,   {Longo} G.,  2015, Exp. Astron., 39, 45

\bibitem[\protect\citeauthoryear{{Chambers, K. C., et al}}{{Chambers, K. C., et al}}{2019}]{chambers2019panstarrs1}
{Chambers, K. C., et al} 2019, The Pan-STARRS1 Surveys

\bibitem[\protect\citeauthoryear{{Connolly}, {Csabai}, {Szalay}, {Koo}, {Kron}  \& {Munn}}{{Connolly} et~al.}{1995}]{1995AJ....110.2655C}
{Connolly} A.~J.,  {Csabai} I.,  {Szalay} A.~S.,  {Koo} D.~C.,  {Kron} R.~G.,   {Munn} J.~A.,  1995, \aj, 110, 2655

\bibitem[\protect\citeauthoryear{Cramér}{Cramér}{1928}]{99}
Cramér H.,  1928, Scand. Actuar. J., 1928, 13

\bibitem[\protect\citeauthoryear{Curran, Moss  \& Perrott}{Curran et~al.}{2021}]{ml4}
Curran S.~J.,  Moss J.~P.,   Perrott Y.~C.,  2021, MRNAS, 503, 2639

\bibitem[\protect\citeauthoryear{{D\'{}Isanto, A.} \& {Polsterer, K. L.}}{{D\'{}Isanto, A.} \& {Polsterer, K. L.}}{2018}]{concave}
{D\'{}Isanto, A.} {Polsterer, K. L.} 2018, A\&A, 609

\bibitem[\protect\citeauthoryear{Dahlen, Mobasher  \& Faber}{Dahlen et~al.}{2013}]{Dahlen_2013}
Dahlen T.,  Mobasher B.,   Faber S. M. e.~a.,  2013, Astrophys. J., 775, 93

\bibitem[\protect\citeauthoryear{{Dey, A., et al}}{{Dey, A., et al}}{2019}]{102}
{Dey, A., et al} 2019, AAS, 157

\bibitem[\protect\citeauthoryear{Dey, Zhao, Andrews, Newman, Izbicki  \& Lee}{Dey et~al.}{}]{marg_cal}
Dey B.,  Zhao D.,  Andrews B.,  Newman J.,  Izbicki R.,   Lee A., , Machine Learning for Astrophysics, proceedings of the Thirty-ninth International Conference on Machine Learning (ICML 2022)

\bibitem[\protect\citeauthoryear{{Dey}, {Andrews}, {Newman}, {Mao}, {Rau}  \& {Zhou}}{{Dey} et~al.}{2022}]{2022MNRAS.515.5285D}
{Dey} B.,  {Andrews} B.~H.,  {Newman} J.~A.,  {Mao} Y.-Y.,  {Rau} M.~M.,   {Zhou} R.,  2022, \mn@doi [\mnras] {10.1093/mnras/stac2105}, \href {https://ui.adsabs.harvard.edu/abs/2022MNRAS.515.5285D} {515, 5285}

\bibitem[\protect\citeauthoryear{Dharani, Nair, Satpathy  \& Christopher}{Dharani et~al.}{2019}]{8978471}
Dharani Y.~G.,  Nair N.~G.,  Satpathy P.,   Christopher J.,  2019, in 2019 Global Conference for Advancement in Technology (GCAT). pp~1--6, \mn@doi{10.1109/GCAT47503.2019.8978471}

\bibitem[\protect\citeauthoryear{{Driver, S.P., et al.}}{{Driver, S.P., et al.}}{2011}]{gama}
{Driver, S.P., et al.} 2011, MNRAS, 413, 971

\bibitem[\protect\citeauthoryear{Duncan et~al.,}{Duncan et~al.}{2017}]{10.1093/mnras/stx2536}
Duncan K.~J.,  et~al., 2017, Monthly Notices of the Royal Astronomical Society, 473, 2655

\bibitem[\protect\citeauthoryear{{Euclid Collaboration} et~al.,}{{Euclid Collaboration} et~al.}{2020}]{refId0}
{Euclid Collaboration} et~al., 2020, A\&A, 644, A31

\bibitem[\protect\citeauthoryear{{Fernández-Soto, A., et al}}{{Fernández-Soto, A., et al}}{1999}]{LY}
{Fernández-Soto, A., et al} 1999, APJ, 513

\bibitem[\protect\citeauthoryear{{Hatfield}, {Jarvis}, {Adams}, {Bowler}, {H{\"a}u{\ss}ler}  \& {Duncan}}{{Hatfield} et~al.}{2022}]{2022MNRAS.513.3719H}
{Hatfield} P.~W.,  {Jarvis} M.~J.,  {Adams} N.,  {Bowler} R.~A.~A.,  {H{\"a}u{\ss}ler} B.,   {Duncan} K.~J.,  2022, \mnras, 513, 3719

\bibitem[\protect\citeauthoryear{{Henghes, B. et al}}{{Henghes, B. et al}}{2022}]{metric}
{Henghes, B. et al} 2022, MNRAS, 512, 1696

\bibitem[\protect\citeauthoryear{Henghes, Pettitt, Thiyagalingam, Hey  \& Lahav}{Henghes et~al.}{2021}]{10.1093/mnras/stab1513}
Henghes B.,  Pettitt C.,  Thiyagalingam J.,  Hey T.,   Lahav O.,  2021, MNRAS, 505, 4847

\bibitem[\protect\citeauthoryear{{Hildebrandt} et~al.,}{{Hildebrandt} et~al.}{2010}]{2010A&A...523A..31H}
{Hildebrandt} H.,  et~al., 2010, \mn@doi [\aap] {10.1051/0004-6361/201014885}, \href {https://ui.adsabs.harvard.edu/abs/2010A&A...523A..31H} {523, A31}

\bibitem[\protect\citeauthoryear{Hoyle}{Hoyle}{2016}]{ml5}
Hoyle B.,  2016, Astron. Comput., 16, 34

\bibitem[\protect\citeauthoryear{{Ilbert, O., et al.}}{{Ilbert, O., et al.}}{2006}]{0.15_ref}
{Ilbert, O., et al.} 2006, A\&A, 457, 841

\bibitem[\protect\citeauthoryear{{Ilbert} et~al.,}{{Ilbert} et~al.}{2006}]{2006A&A...457..841I}
{Ilbert} O.,  et~al., 2006, \aap, 457, 841

\bibitem[\protect\citeauthoryear{Janiurek}{Janiurek}{2023}]{me}
Janiurek L.,  2023, Master's thesis, University of Glasgow, https://theses.gla.ac.uk/83369/

\bibitem[\protect\citeauthoryear{{Jarvis} et~al.,}{{Jarvis} et~al.}{2013}]{2013MNRAS.428.1281J}
{Jarvis} M.~J.,  et~al., 2013, \mnras, 428, 1281

\bibitem[\protect\citeauthoryear{Jones}{Jones}{2022}]{jones2023photometric}
Jones E. e.~a.,  2022, in NeurIPS 2021.

\bibitem[\protect\citeauthoryear{Jones \& Heavens}{Jones \& Heavens}{2018}]{10.1093/mnras/sty3279}
Jones D.~M.,  Heavens A.~F.,  2018, MNRAS, 483, 2487

\bibitem[\protect\citeauthoryear{{Jones} \& {Singal}}{{Jones} \& {Singal}}{2017}]{2017A&A...600A.113J}
{Jones} E.,  {Singal} J.,  2017, \mn@doi [\aap] {10.1051/0004-6361/201629558}, \href {https://ui.adsabs.harvard.edu/abs/2017A&A...600A.113J} {600, A113}

\bibitem[\protect\citeauthoryear{Kind \& Brunner}{Kind \& Brunner}{2013}]{82}
Kind M.,  Brunner R.,  2013, MNRAS, 432, 1483

\bibitem[\protect\citeauthoryear{Kullback \& Leibler}{Kullback \& Leibler}{1951}]{98}
Kullback R.,  Leibler S.,  1951, Ann. Math. Statist., 22, 79

\bibitem[\protect\citeauthoryear{Lanzetta, Yahil  \& Fernandez-Soto}{Lanzetta et~al.}{1998}]{LYF}
Lanzetta K.,  Yahil A.,   Fernandez-Soto A.,  1998, AAS, 116, 1066

\bibitem[\protect\citeauthoryear{Lee}{Lee}{2024}]{particles7020019}
Lee S.,  2024, Particles, 7, 309

\bibitem[\protect\citeauthoryear{{Lee, K.G., et al.}}{{Lee, K.G., et al.}}{2013}]{boss}
{Lee, K.G., et al.} 2013, AJ, 145, 69

\bibitem[\protect\citeauthoryear{Leistedt, Alsing, Peiris, Mortlock  \& Leja}{Leistedt et~al.}{2023}]{Leistedt_2023}
Leistedt B.,  Alsing J.,  Peiris H.,  Mortlock D.,   Leja J.,  2023, The Astrophysical Journal Supplement Series, 264, 23

\bibitem[\protect\citeauthoryear{Lu et~al.,}{Lu et~al.}{2023}]{10.1093/mnras/stad3976}
Lu J.,  et~al., 2023, Monthly Notices of the Royal Astronomical Society, 527, 12140

\bibitem[\protect\citeauthoryear{{Lupton}, {Gunn}  \& {Szalay}}{{Lupton} et~al.}{1999}]{1999AJ....118.1406L}
{Lupton} R.~H.,  {Gunn} J.~E.,   {Szalay} A.~S.,  1999, \aj, 118, 1406

\bibitem[\protect\citeauthoryear{{Marocco, F., et al}}{{Marocco, F., et al}}{2021}]{89}
{Marocco, F., et al} 2021, ApJS, 235

\bibitem[\protect\citeauthoryear{{Massimo}}{{Massimo}}{2021}]{Brescia_2021}
{Massimo} B. e.~a.,  2021, Front. astron. space sci, 8

\bibitem[\protect\citeauthoryear{{McCracken} et~al.,}{{McCracken} et~al.}{2012}]{2012A&A...544A.156M}
{McCracken} H.~J.,  et~al., 2012, \aap, 544, A156

\bibitem[\protect\citeauthoryear{{Momtaz, A. et al.}}{{Momtaz, A. et al.}}{2022}]{b1}
{Momtaz, A. et al.} 2022, {Estimating the photometric redshifts of galaxies and QSOs using regression techniques in machine learning}.
pp 368--381

\bibitem[\protect\citeauthoryear{Mucesh}{Mucesh}{2021}]{galpro-github}
Mucesh S.,  2021, MNRAS, 502, 2770

\bibitem[\protect\citeauthoryear{{Mucesh} et~al.,}{{Mucesh} et~al.}{2021}]{2021MNRAS.502.2770M}
{Mucesh} S.,  et~al., 2021, \mn@doi [\mnras] {10.1093/mnras/stab164}, \href {https://ui.adsabs.harvard.edu/abs/2021MNRAS.502.2770M} {502, 2770}

\bibitem[\protect\citeauthoryear{{Myles, J. et al}}{{Myles, J. et al}}{2022}]{pit15}
{Myles, J. et al} 2022, MNRAS, 519, 1792

\bibitem[\protect\citeauthoryear{Newman \& Gruen}{Newman \& Gruen}{2022}]{red2}
Newman J.~A.,  Gruen D.,  2022, Annual Review of Astronomy and Astrophysics, 60, 363

\bibitem[\protect\citeauthoryear{Norris et~al.,}{Norris et~al.}{2019}]{Norris_2019}
Norris R.~P.,  et~al., 2019, Publications of the Astronomical Society of the Pacific, 131, 108004

\bibitem[\protect\citeauthoryear{O. \& S.}{O. \& S.}{2003}]{num1}
O. F. A. E.~L.,  S. S.~R.,  2003, MNRAS, 339, 1195

\bibitem[\protect\citeauthoryear{Quandri \& Williams}{Quandri \& Williams}{2010}]{Quadri_2010}
Quandri R.~F.,  Williams R.~J.,  2010, AJ, 725, 794

\bibitem[\protect\citeauthoryear{Rasmussen \& Williams}{Rasmussen \& Williams}{2006}]{RasmussenW06}
Rasmussen C.~E.,  Williams C. K.~I.,  2006, Gaussian processes for machine learning..
Adaptive computation and machine learning, MIT Press

\bibitem[\protect\citeauthoryear{Razim, Cavuoti, Brescia, Riccio, Salvato  \& Longo}{Razim et~al.}{2021}]{ml2}
Razim O.,  Cavuoti S.,  Brescia M.,  Riccio G.,  Salvato M.,   Longo G.,  2021, MNRAS, 507, 5034

\bibitem[\protect\citeauthoryear{{SDSS-III Collaboration}}{{SDSS-III Collaboration}}{2012}]{100}
{SDSS-III Collaboration} 2012, AAS, 203, 21

\bibitem[\protect\citeauthoryear{Sadeh, Abdalla  \& Lahav}{Sadeh et~al.}{2016}]{ml3}
Sadeh I.,  Abdalla F.,   Lahav O.,  2016, Publications of the Astronomical Society of the Pacific, 128, 104502

\bibitem[\protect\citeauthoryear{Salvato, Ilbert  \& Hoyle}{Salvato et~al.}{2018}]{flavours}
Salvato M.,  Ilbert O.,   Hoyle B.,  2018, Nat Ast, 3, 212

\bibitem[\protect\citeauthoryear{{Schmidt, S.J., et al.}}{{Schmidt, S.J., et al.}}{2020}]{90}
{Schmidt, S.J., et al.} 2020, MNRAS, 499, 1587

\bibitem[\protect\citeauthoryear{Steinhardt, Kokorev, Rusakov, Garcia  \& Sneppen}{Steinhardt et~al.}{2023}]{Steinhardt_2023}
Steinhardt C.~L.,  Kokorev V.,  Rusakov V.,  Garcia E.,   Sneppen A.,  2023, The Astrophysical Journal Letters, 951, L40

\bibitem[\protect\citeauthoryear{Stephens}{Stephens}{1992}]{97}
Stephens M.,  1992, Springer New York, pp 93--105

\bibitem[\protect\citeauthoryear{Sánchez et~al.,}{Sánchez et~al.}{2014}]{Sanchez}
Sánchez C.,  et~al., 2014, MNRAS, 445, 1482

\bibitem[\protect\citeauthoryear{Tanaka}{Tanaka}{2015}]{Tanaka_2015}
Tanaka M.,  2015, Astrophys. J., 801, 20

\bibitem[\protect\citeauthoryear{{Tanigawa}, {Glazebrook}, {Jacobs}, {Labbe}  \& {Qin}}{{Tanigawa} et~al.}{2024}]{2024MNRAS.530.2012T}
{Tanigawa} S.,  {Glazebrook} K.,  {Jacobs} C.,  {Labbe} I.,   {Qin} A.~K.,  2024, \mn@doi [\mnras] {10.1093/mnras/stae411}, \href {https://ui.adsabs.harvard.edu/abs/2024MNRAS.530.2012T} {530, 2012}

\bibitem[\protect\citeauthoryear{Trinchera}{Trinchera}{2023}]{space}
Trinchera A.,  2023, Front. Astron. Space Sci., 9

\bibitem[\protect\citeauthoryear{Walcher \& T.~Dale}{Walcher \& T.~Dale}{2010}]{red45}
Walcher J.~Groves B.~B.,  T.~Dale D.,  2010, Ap\&SS, 331, 1

\bibitem[\protect\citeauthoryear{Wittman}{Wittman}{2009}]{Wittman_2009}
Wittman D.,  2009, AJ, 700, L174

\bibitem[\protect\citeauthoryear{Wolf}{Wolf}{2009}]{wolf}
Wolf C.,  2009, MRNAS, 397, 520

\bibitem[\protect\citeauthoryear{Zhou}{Zhou}{2021}]{zhou-desi}
Zhou R.,  2021, MNRAS, 501, 3309

\bibitem[\protect\citeauthoryear{Zhou et~al.,}{Zhou et~al.}{2021}]{Zhou_2021}
Zhou X.,  et~al., 2021, \mn@doi [The Astrophysical Journal] {10.3847/1538-4357/abda3e}, 909, 53

\makeatother
\end{thebibliography}




\appendix

\section{Sanity Check Results}\label{sane}

This appendix presents the $r$-band magnitude distributions and GALPRO output plots for the two sanity check tests described in Section \ref{sancheck}, where the $r$-band distributions of the testing and training samples drawn from the Zhou et al. dataset are chosen to overlap by either 100\% or 0\%, representing the extreme cases whereby the two 'surveys' are completely statistical equivalent or have no statistical equivalence. 

\subsection{100\% Overlap Test}

\begin{figure}
  \centering

  \begin{tabular}{@{}c@{}}
    \includegraphics[width=\columnwidth]{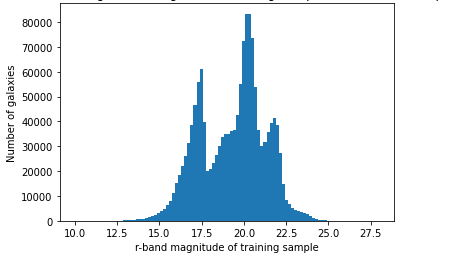} \\[\abovecaptionskip]
    \small (a) $r$-band distribution of training data (Survey 1).
  \end{tabular}

    \vspace{\floatsep}

    \centering
  \begin{tabular}{@{}c@{}}
    \includegraphics[width=\columnwidth]{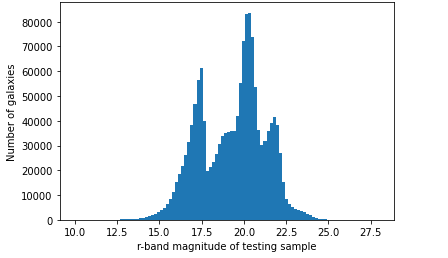} \\[\abovecaptionskip]
    \small (b) $r$-band distribution of testing data (Survey 2).
  \end{tabular}
  
  \caption{$r$-band magnitude distributions of the training data (Survey 1) and testing data (Survey 2), for the case with 100\% overlap between these two surveys.}\label{appen_100}
\end{figure}

First, we show the results of the 100\% overlap test, whereby the testing and training samples drawn from the Zhou et al. datset and inputted to GALPRO overlap in $r$-band magnitude by 100\%. Figures \ref{appen_100} show the $r$-band magnitude distributions of the testing and training samples in this case. It is clear from the plots that the $r$-band distributions of the two samples are identical. 

\begin{figure}
  \centering
  
  \begin{tabular}{ c c }
    \includegraphics[width=0.65\columnwidth]{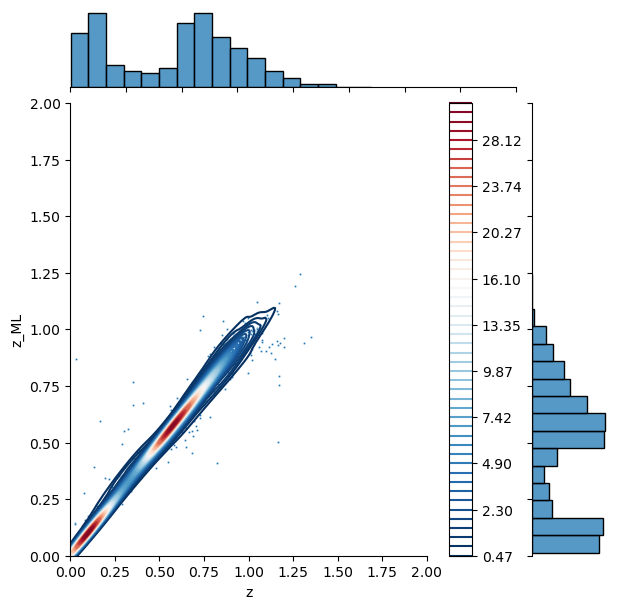} \\[\abovecaptionskip]
    \small (a) Photometric versus spectroscopic redshift plot.
  \end{tabular}

  \vspace{\floatsep}

   \centering
  \begin{tabular}{ c c }
    \includegraphics[width=0.65\columnwidth]{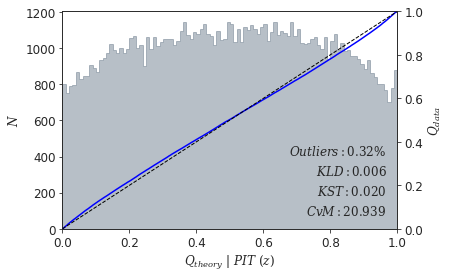} \\[\abovecaptionskip]
    \small (b) PIT plot, also showing the outlier fraction,\\ 
    KLD, KST and CvM test statistics.
  \end{tabular}

  \vspace{\floatsep}

   \centering
  \begin{tabular}{ c c }
    \includegraphics[width=0.65\columnwidth]{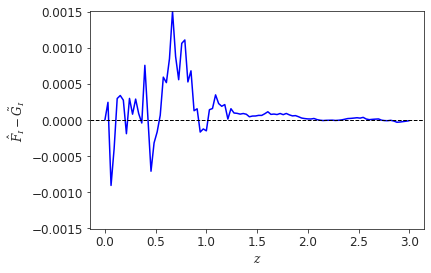} \\[\abovecaptionskip]
    \small (c) Marginal calibration plot.
  \end{tabular}

  \caption{GALPRO results when trained using the Zhou et al. sample and adopting a 100\% overlap between the training and testing datasets. In the upper panel the dots show the photometric redshift estimate (vertical axis) versus the true spectroscopic redshift (horizontal axis) for every galaxy in the testing sample. Also plotted are coloured contours representing the joint likelihood of the photometric and spectroscopic redshifts, with the contour colour bar  displaying log likelihood values. Above and to the right of the plot are histograms of the redshift distributions for the training and testing samples respectively. The middle panel plots the PIT for the calibration sample and also shows the outlier fraction and the values of the KLD, KST and CvM test statistics. The lower panel shows the marginal calibration plot for the calibration sample.}\label{100_res}
\end{figure}

Figures \ref{100_res} then show the spectroscopic versus photometric redshift plot and the PIT and marginal calibration plots outputted by GALPRO when these samples are used as inputs. As expected, the plots show that the photometric redshift PDFs are valid and have been successfully calibrated. 

\subsection{0\% Overlap Test}\label{0_appen}

\begin{figure}
  \centering

  \begin{tabular}{@{}c@{}}
    \includegraphics[width=\columnwidth]{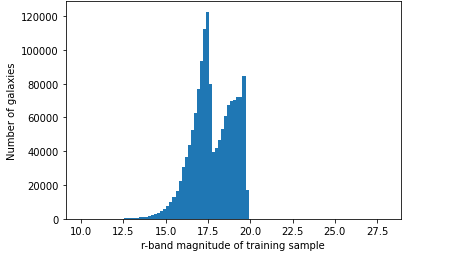} \\[\abovecaptionskip]
    \small (a) $r$-band distribution of training data (Survey 1).
  \end{tabular}

\vspace{\floatsep}

   \centering
  \begin{tabular}{@{}c@{}}
    \includegraphics[width=\columnwidth]{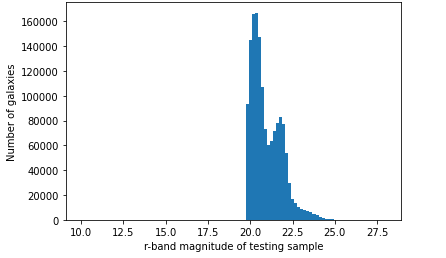} \\[\abovecaptionskip]
    \small (b) $r$-band distribution of testing data (Survey 2).
  \end{tabular}
  
  \caption{$r$-band magnitude distributions of the training data (Survey 1) and testing data (Survey 2), for the case with 0\% overlap between these two surveys.}\label{appn_0}
\end{figure}

Next we show the results of the  0\% overlap test, whereby the testing and training samples inputted to GALPRO have no overlap in $r$-band magnitude. Figures \ref{appn_0} show the $r$-band magnitude distributions of the testing and training samples in this case. It is clear from these plots that the $r$-band distributions of the two samples are completely disjoint.

\begin{figure}
  \centering
  
  \begin{tabular}{ c c }
    \includegraphics[width=0.65\columnwidth]{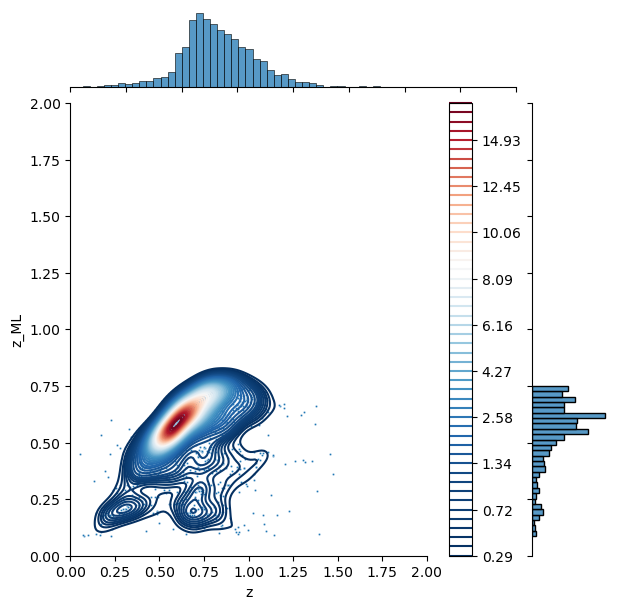} \\[\abovecaptionskip]
    \small (a) Photometric versus spectroscopic redshift plot.
  \end{tabular}

  \vspace{\floatsep}

   \centering
  \begin{tabular}{ c c }
    \includegraphics[width=0.65\columnwidth]{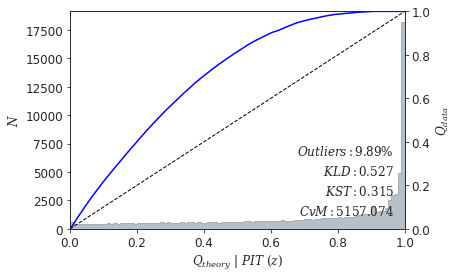} \\[\abovecaptionskip]
    \small (b) PIT plot, also showing the outlier fraction,\\ 
    KLD, KST and CvM test statistics.
  \end{tabular}

  \vspace{\floatsep}
  
   \centering
  \begin{tabular}{ c c }
    \includegraphics[width=0.65\columnwidth]{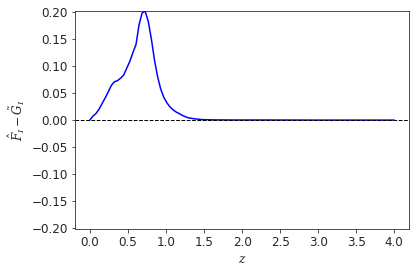} \\[\abovecaptionskip]
    \small (c) Marginal calibration plot.
  \end{tabular}

  \caption{GALPRO results when trained using the Zhou et al. sample and adopting a 0\% overlap between the training and testing datasets. In the upper panel the dots show the photometric redshift estimate (vertical axis) versus the true spectroscopic redshift (horizontal axis) for every galaxy in the testing sample. Also plotted are coloured contours representing the joint likelihood of the photometric and spectroscopic redshifts, with the contour colour bar  displaying log likelihood values. Above and to the right of the plot are histograms of the redshift distributions for the training and testing samples respectively. The middle panel plots the PIT for the calibration sample and also shows the outlier fraction and the values of the KLD, KST and CvM test statistics. The lower panel shows the marginal calibration plot for the calibration sample.}\label{0_res}
\end{figure}

Figures \ref{0_res} then show the spectroscopic versus photometric redshift plot and the PIT and marginal calibration plots output by GALPRO when these samples are used as inputs. As expected, the plots show that the photometric redshift PDFs are highly inaccurate and indicate an unsuccessful calibration. 

\section{The $r$-band magnitude distributions used in the overlap tests}\label{rtestsss}

\begin{figure}
  \centering

  \begin{tabular}{@{}c@{}}
    \includegraphics[width=\columnwidth]{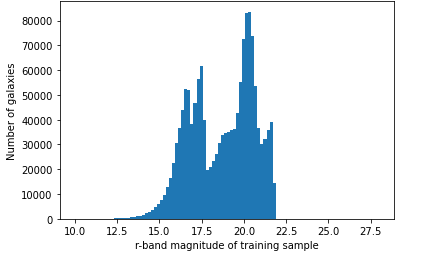} \\[\abovecaptionskip]
    \small (a) $r$-band distribution of training data (Survey 1).
  \end{tabular}

    \vspace{\floatsep}

  \begin{tabular}{@{}c@{}}
    \includegraphics[width=\columnwidth]{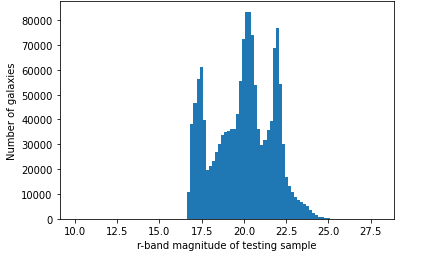} \\[\abovecaptionskip]
    \small (b) $r$-band distribution of testing data (Survey 2).
  \end{tabular}
  
  \caption{$r$-band magnitude distributions of the training data (Survey 1) and testing data (Survey 2), for the case with 90\% overlap between these two surveys.}\label{rband90}
\end{figure}

\begin{figure}
  \centering

    \begin{tabular}{@{}c@{}}
    \includegraphics[width=\columnwidth]{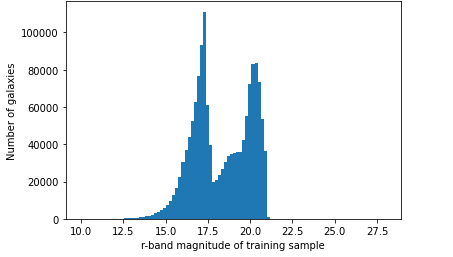} \\[\abovecaptionskip]
    \small (a) $r$-band distribution of training data (Survey 1).
  \end{tabular}

     \vspace{\floatsep}

  \begin{tabular}{@{}c@{}}
    \includegraphics[width=\columnwidth]{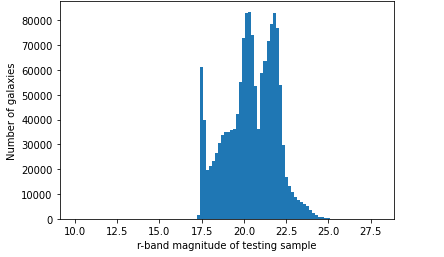} \\[\abovecaptionskip]
    \small (b) $r$-band distribution of testing data (Survey 2).
  \end{tabular}
  
  \caption{$r$-band magnitude distributions of the training data (Survey 1) and testing data (Survey 2), for the case with 80\% overlap between these two surveys.}\label{rband80}
\end{figure}

\begin{figure}
  \centering

  \begin{tabular}{@{}c@{}}
    \includegraphics[width=\columnwidth]{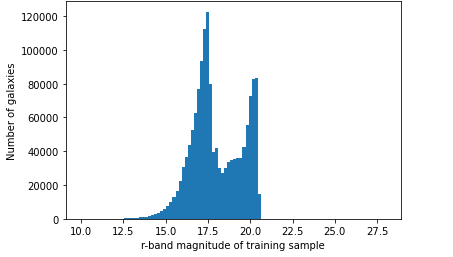} \\[\abovecaptionskip]
    \small (a) $r$-band distribution of training data (Survey 1).
  \end{tabular}

    \vspace{\floatsep}

  \begin{tabular}{@{}c@{}}
    \includegraphics[width=\columnwidth]{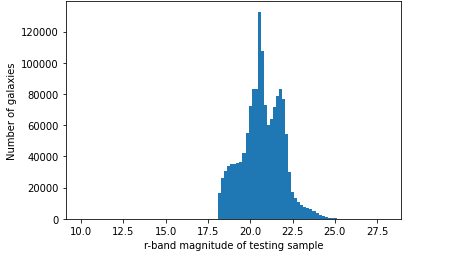} \\[\abovecaptionskip]
    \small (b) $r$-band distribution of testing data (Survey 2).
  \end{tabular}
  
  \caption{$r$-band magnitude distributions of the training data (Survey 1) and testing data (Survey 2), for the case with 70\% overlap between these two surveys.}\label{rband70}
\end{figure}

This appendix presents the $r$-band magnitude distributions of the training and testing samples inputted to GALPRO, as used in the overlap tests described in Section \ref{overlap}. The $r$-band distributions of the testing and training samples are shown for the cases where the distributions overlap by 90\% (Figure \ref{rband90}), 80\% (Figure \ref{rband80}) and 70\% (Figure \ref{rband70}) respectively.

\bsp	
\label{lastpage}
\end{document}